%% file: main.tex
\documentclass{article}
\usepackage{hyperref}
\hypersetup{
    colorlinks=false,
    linkcolor=black,
    filecolor=magenta,      
    urlcolor=black,
}
\usepackage[utf8]{inputenc}
\usepackage{graphicx}
\usepackage[nottoc]{tocbibind}

\usepackage{ltablex}
\usepackage{multirow} 

\usepackage[a4paper, total={6in, 8in}]{geometry}

\usepackage{colortbl}

\title{Bibliography management: BibTeX}
\author{Share\LaTeX}

\begin{document}

\input{title}
\newpage
\input{aboutIVI}

\newpage
\tableofcontents
\newpage
\input{summary}

\newpage
\input{introduction}
\newpage
\input{abstractionLevels}
\input{methodology}

\newpage
\input{assessmentFramework}

\newpage
\input{irelandAssessment}

\newpage
\input{lessons_learned}

\input{conclusion}
\newpage
\input{bios}

\bibliographystyle{apalike}
\bibliography{references}

\end{document}

%% file: title.tex
\begin{titlepage}

\newcommand{\HRule}{\rule{\linewidth}{0.5mm}} 

\center 
 

\textsc{\LARGE Innovation Value Institute}\\[1.5cm] 
\includegraphics[scale=.2]{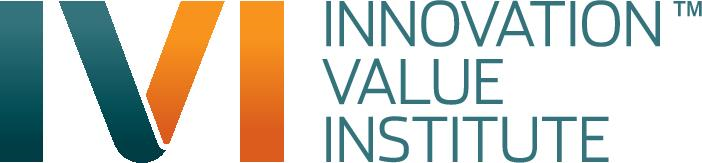}\\[1cm] 


\HRule \\[0.4cm]
{ \huge \bfseries An Assessment Methodology and Instrument for Cybersecurity: The Ireland Use Case}\\[0.4cm] 
\HRule \\[1.5cm]
\textsc{\Large White Paper}\\[0.5cm] 



\begin{minipage}{.6\textwidth}
\begin{center} \large
Marco Alfano, Viviana Bastidas, \\ Paul Heynen, Markus Helfert\\ 
\end{center}

\end{minipage}\\[2cm]



{\large \today}\\[2cm] 

\vfill 

\end{titlepage}

%% file: aboutIVI.tex
\addcontentsline{toc}{section}{About Innovation Value Institute (IVI)}
\section*{About Innovation Value Institute (IVI)}

IVI\footnote{Innovation Value Institute: \url{https://ivi.ie/}} is a multidisciplinary research institute focused on digital transformation and technology adoption based in Maynooth University.
Founded in 2006 in collaboration with Intel, we have a strong track record of industry collaboration, both locally and internationally.
We have demonstrated excellent dissemination capability including education and training, and have developed a close working relationship with Enterprise Ireland, IDA Ireland, Science Foundation Ireland and other research centres. Research is focused on discovering strategic research challenges and co-creating validated Digital Transformation Paths. Together with the research partners in the Research Clusters, IVI develops innovative research outputs that are relevant for both research and in practice, to transform and architect sustainable Service Ecosystems.
 
In particular, The IT Capability Maturity Framework (IT-CMF) is a framework specifically created to derive real, measurable business value from IT. It helps organisations devise more robust strategies, make better-informed decisions, and perform more effectively, efficiently and consistently. IT-CMF acts as a unifying (or umbrella) framework that complements other domain-specific frameworks already in use in the organization, helping to resolve conflicts between them, and filling gaps in their coverage. It provides a holistic, business-led approach that is:

\begin{itemize}

    \item \textbf{Helpful:} IT-CMF helps organizations to develop enduring IT capabilities. 
    
    \item \textbf{Coherent:} It is underpinned by coherent concepts and principles that help stakeholders to agree strategic goals, implement planned actions and evaluate performance.
    
    \item \textbf{Complementary:} It complements other, domain-specific frameworks already in use in the organization and fills in gaps in coverage.

    \item \textbf{Scaleable:} It can be used to guide performance improvement in organizations of any size and in any sector.

\end{itemize}

%% file: summary.tex
\addcontentsline{toc}{section}{Summary}
\section*{Summary}

Governments around the world are required to strengthen their national cybersecurity capabilities to respond effectively to the growing, changing, and sophisticated cyber threats and attacks, thus protecting society and the way of life as a whole. Responsible government institutions need to revise, evaluate, and bolster their national cybersecurity capabilities to fulfill the new requirements, for example regarding new trends affecting cybersecurity, key supporting laws and regulations, and implementations risk and challenges. This report presents a comprehensive assessment instrument for cybersecurity at the national level in order to help countries to ensure optimum response capability and more effective use of critical resources of each state. More precisely, the report

\begin{itemize}

    \item builds a common understanding of the critical cybersecurity capabilities and competence to be assessed at the national level, 
    
    \item adds value to national strategic planning and implementation which impact the development and adaptation of national cybersecurity strategies,
    
    \item provides an overview of the assessment approaches at the national level, including capabilities, frameworks, and controls, 

    \item introduces a comprehensive cybersecurity instrument for countries to determine areas of improvement and develop enduring national capabilities,
    
    \item describes how to apply the proposed national cybersecurity assessment framework in a real-world case, and
    
    \item presents the results and lessons learned of the application of the assessment framework at the national level to assist governments in further building cybersecurity capabilities.
    
\end{itemize}

%% file: introduction.tex
\section{Introduction: National Cybersecurity Assessment}

Several countries have recognised that cybersecurity is a key national security priority. Government institutions are exposed to an increased number of cyberthreats and new types of attacks. These include, for example, recent waves of ransomware attacks on national critical systems compromising information and telecommunication technology networks and the delivery of vital public services~\cite{sadik2020toward}. For example, in 2021, the Health Service Executive (HSE) of Ireland  suffered a major ransomware cyberattack that caused all its IT systems nationwide to be shut down. According to the report of the HSE\footnote{HSE lessons learned report: \url{https://www.hhs.gov/sites/default/files/lessons-learned-hse-attack.pdf}}, 80\% of the HSE IT environment was encrypted, severely disrupting healthcare services throughout the country. This year, Royal Mail Ltd, a British multinational postal service and courier company, was the victim of a ransomware attack this year, and it is still working with security authorities to understand and mitigate the impact\footnote{NCSC statement on the Royal Mail incident: \url{https://www.ncsc.gov.uk/news/royal-mail-incident}}. In such cases, ransomware victims may not only incur economic losses, but they may also suffer other damages such as loss of sensitive data and reputation~\cite{al2018ransomware}. Given these situations, government leaders have assessed the emerging risks from technological adoption, which materialised even more with the transformation forced by the COVID-19 pandemic~\cite{chigada2021cyberattacks}.

Over the past years, several countries have adopted different actions to address these challenges posed by cyberthreats. These actions involve, among others, the definition of cybersecurity laws and regulations, the constitution of national cybersecurity committees, the development of national cybersecurity strategies, and the creation of cybersecurity implementation frameworks~\cite{sabillon2016national}. In particular, national cybersecurity strategies set the goals, objectives, and the course for national efforts to strengthen cybersecurity~\cite{dedeke2019contrasting}. On the other hand, cybersecurity implementation frameworks provide a guide for the execution of national cybersecurity strategies and agendas. Such frameworks are mainly proposed by international standards organisations (ISO/IEC 27000 series), by private industry-related initiatives (e.g., COBIT), and by government-led initiatives which involve public-private partnerships (e.g., National Institute of Standards and Technology – NIST Framework). These implementation frameworks focus mostly on essential requirements, practices, processes, and controls for implementing information security management systems. However, they pay less attention to the abilities and resources needed to develop national capabilities regarding cybersecurity.


Capability-based frameworks for national cybersecurity should define the abilities that government institutions must have for the implementation of strategic goals. Countries must adapt to the dynamic nature of these capabilities and respond to the rapid change in the cybersecurity threat landscape. According to the report after the NHS case, reducing cybersecurity risk requires both a transformation in cybersecurity capability and IT transformation, to address the issues of a legacy IT estate and build cybersecurity and resilience into the IT architecture~\cite{web:hse2021report}. In this direction, the European Union Agency for Cybersecurity (ENISA) proposes the National Capabilities Assessment Framework (NCAF)~\cite{enisa2020frame}. The framework aims to provide an assessment of the national level of maturity by evaluating strategic objectives. Despite these international efforts, there is a need for a comprehensive assessment instrument that enables the evaluation and development of a broader set of national cybersecurity capabilities. This instrument should help government institutions to identify cybersecurity gaps and limitations and provide the best practices to drive improvement.

This report presents a comprehensive assessment instrument for cybersecurity at the national level which is presented as a capability-based framework. The proposed framework aims to help countries to ensure optimum response capability and more effective use of critical resources. It is structured around four main capability categories and twenty capability building blocks:

\begin{itemize}
    \item  \textbf{Capability Category: Governance}
        \begin{itemize}
            \item[] 1. Information Security Principles, Policies, and Controls
            \item[] 2. Information Security Strategy
            \item[] 3. Governance Structure
            \item[] 4. Roles, Responsibilities, and Accountabilities
            \item[] 5. Security Risk Management
            \item[] 6. Skills and Competence Development
            \item[] 7. Security Performance Measurement
            \item[] 8. Cybersecurity Implementation Frameworks
            \item[] 9. Cybersecurity Implementations Risks and Challenges
            \item[] 10. Cybersecurity Technology Trends
        \end{itemize}
    \item \textbf{Capability Category: Technical Security}
        \begin{itemize}
            \item[] 11. IT Device Security
            \item[] 12. Cybersecurity Threats
            \item[] 13. Cybersecurity Threat Actors
        \end{itemize}
    \item \textbf{Capability Category: Security Data Administration}
        \begin{itemize}
            \item[] 15. Data Security Classification
            \item[] 16. Data Life Cycle Management
            \item[] 17. Data Security Administration
            \item[] 18. Identity and Access Management
        \end{itemize}
    \item \textbf{Capability Category: Business Continuity Management}
        \begin{itemize}
            \item[] 19. Business Continuity Planning and Management
            \item[] 20. Cybersecurity Threat Actors
        \end{itemize}
\end{itemize}



%% file: abstractionLevels.tex
\section{Cybersecurity Abstraction Levels}

\begin{figure}[!ht]
	\centering
	\includegraphics[width=0.95\linewidth]{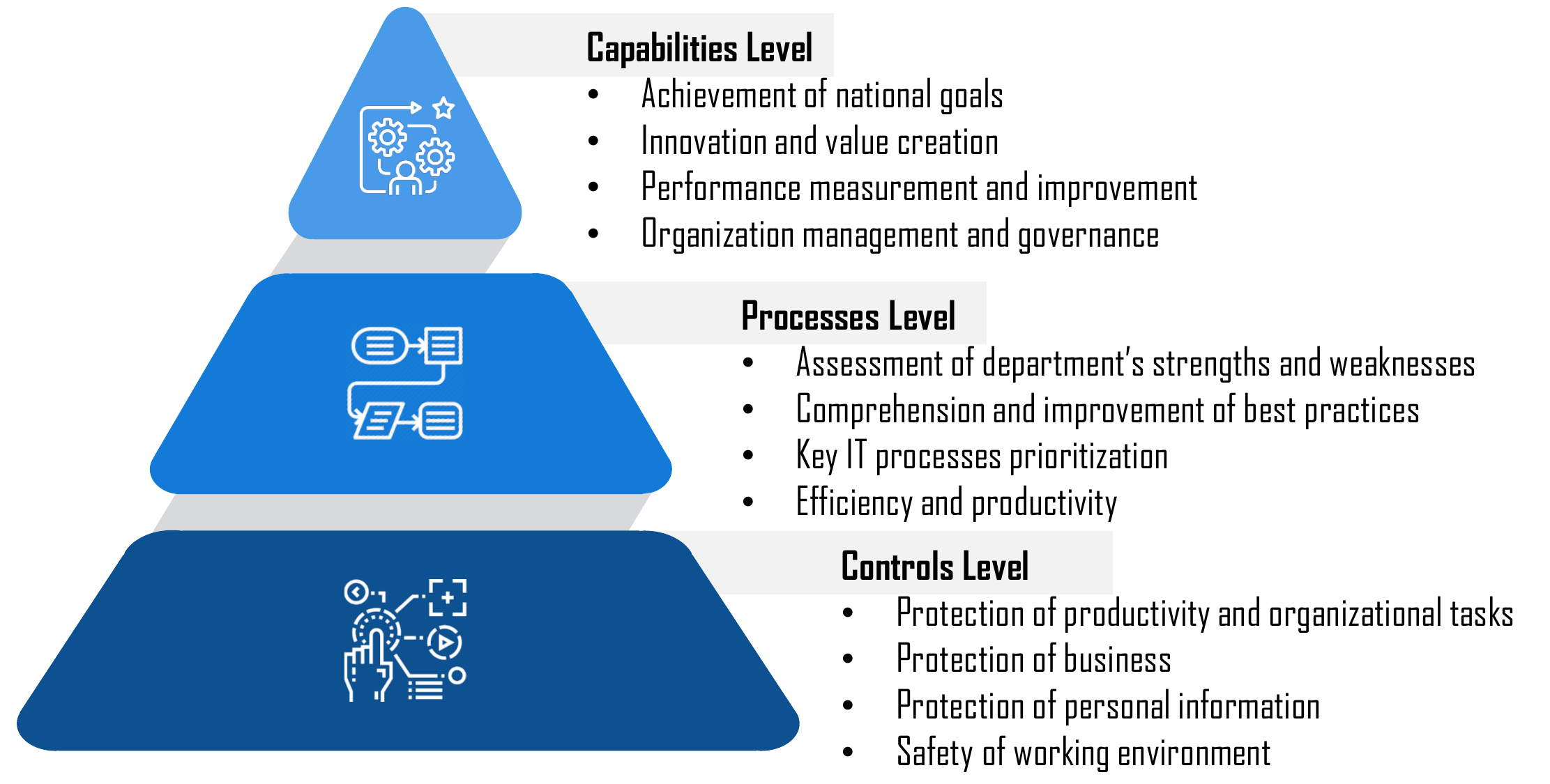}
	\caption{Cybersecurity Abstractions Levels - Authors' work }
\end{figure}


\subsection{Capabilities Level}

This level describes the cybersecurity capabilities as a tool to achieve the national cybersecurity objectives. Such capabilities refer to the abilities required by organisations to implement the main motivations and strategic plans. In particular, cybersecurity capabilities provides the means to evaluate the current situation of cybersecurity and determine where it should stand in comparison. The selection of the appropriate set of cybersecurity capabilities can enable public and private organisations to capture the full promise and value of new technologies while keeping safer places. The evaluation and analysis of these capabilities provides a means to manage and improve the performance of public and private bodies responsible for or involved in the design and implementation of national cybersecurity. Such organisations then can operate effectively and deliver expected outcomes according to their cybersecurity strategies. 

\subsection{Processes Level}

The purpose of this level is to support the implementation of the different capabilities by specifying the associated cybersecurity processes and activities. The definition of different aspects of cybersecurity at this level could help organisations to assess the departments' strengths and weaknesses. At the same time, it can assist people responsible of cybersecurity in the comprehension and improvement of best practices, the prioritization of key IT processes as well as the improvement of efficiency and productivity.

\subsection{Controls Level}

This level outlines the different cybersecurity controls that need to be in place to perform organisational tasks. This level supports the implementation of related organisational processes. Defining different controls helps protect productivity. It provides a means to safeguard the general organization and personal information of individuals in a secure work environment.

%% file: methodology.tex
\section{Methodology}

The structured methodology applied for this study involves four main steps which are outlined below. It is based on sourcing and analysing documents for each country. The document analysis builds the foundation for the qualitative analysis. All documents used with clear references are available on the central repository for subsequent review.

\subsection{Data Gathering Process}

The data gathering process is guided by these detailed steps to ensure objectivity and a balanced document selection process.

\subsubsection{Repositories}

\begin{itemize}
    \item National Cyber Security Index 
    \item Global Cybersecurity Index
    \item Global Cyber Strategies Index
    \item ITU National Cybersecurity Strategies Repositories
    \item Cybersecurity Ranking Repositories
    \item National Cybersecurity Repositories
    \item Online and Academic Repositories
\end{itemize}

\subsubsection{Key Terms}

\begin{itemize}
    \item National cybersecurity strategy  (NCS) + country
    \item Cybersecurity Implementation framework (CIF) + country 
    \item NIST + country framework
    \item Country cybersecurity
    \item Country cybersecurity + IT sector
    \item Cybersecurity index + country
\end{itemize}

\subsubsection{Inclusion and Exclusion Criteria}

\begin{itemize}
    \item Documents had to be a National  cybersecurity strategy 
    \item Documents had to be a Cybersecurity Implementation framework 
    \item Documents had to have been issued by an agency that oversaw national cybersecurity implementation
    \item Document had to be issued by a recognized a cybersecurity institution
\end{itemize}

\subsection{Document analysis method}

\begin{itemize}
  \item Expert panel evaluate sources of evidence to seek convergence and corroboration about the current topic (cybersecurity at the national level).
\end{itemize}

\subsection{Coding process and expert panel agreement}

\begin{itemize}
  \item Assessment Matrix: The main categories of IT-CMF CBB (including questions) that were relevant for each LPL domain and Scope.
  \item List of Documents: The main documents selected that covered the themes in the assessment matrix.
  \item Scoring Agreement: Score how well each of the documents covered the themes that were included in the assessment matrix. 
\end{itemize}

\subsection{Assessment matrix and scoring scheme}

\begin{itemize}
  \item Maturity Score: In scoring each of the documents, a five-point scoring system is adopted according to the maturity levels defined.
\end{itemize}

\subsection{Adding qualitative findings}

\begin{itemize}
  \item Cybersecurity standards/frameworks recommended by government (COBIT, NIST, ISO, etc.).
  \item Cybersecurity per IT Sector (if there is this information at the country level).
\end{itemize}

%% file: assessmentFramework.tex
\section{National Cybersecurity Assessment Framework}

\subsection{IT-CMF}

Organizations, both public and private, are constantly challenged to innovate and generate real value. CIOs are uniquely well-positioned to meet these challenges and to adopt the role of business transformation partner, helping their organizations to grow and prosper with innovative, IT-enabled products, services, and processes. The IT function needs to manage an array of discrete but interdependent disciplines focused on the generation of IT-enabled agility, innovation and value. 

IT Capability Maturity Framework™ (IT-CMF™) is a comprehensive suite of proven management practices, assessment approaches and improvement strategies, developed by the Innovation Value Institute (IVI). It is divided into four Macro-Capabilities:

\begin{itemize}
  \item Managing IT like a Business
  \item Managing the IT Budget
  \item Managing IT for Business Value
  \item Managing the IT Capability
\end{itemize}

and 37 management disciplines or Critical Capabilities (CCs).
For each capability, IT-CMF incorporates a comprehensive suite of maturity profiles, assessment methods, and improvement roadmaps – these are expressed in business language that can be used to guide discussions on setting goals and evaluating performance.

IT-CMF helps organizations devise more robust strategies, make better-informed decisions, and perform more effectively, efficiently and consistently.

In particular, the following Critical Capabilities are related to cybersecurity:
\begin{itemize}
  \item Information Security Management (ISM)
  \item Risk Management (RM)
  \item Personal Data Protection (PDP)
\end{itemize}

Moreover, a specific Cybersecurity Assessment has been developed on this topic  (https://surveys. ivi.ie/+questionnaire/308/+preview).

\subsection{Capability Categories and Building Blocks}
The IT-CMF has been created to evaluate the maturity levels of organization, both public and private. To evaluate the cybersecurity maturity  at the country level, the IT-CMF has been expanded to measure the capability maturity of the practices, standards, policies and frameworks in use in countries.
In particular the Capability Categories and Building Blocks related to cybersecurity have been selected from the above Critical Capabilities and the questions connected to these Building Blocks have been rephrased in order to adapt them to the different cybersecurity aspects of interest for a country.

This has brought to the creation of four Capability Categories and twenty building blocks and related questions as shown in  Table~\ref{tab:ccBbQuestions}.

\footnotesize{
\begin{longtable}{ | m{7.6em} | m{12.4em}| m{8cm} |} 
\caption{Capabilities categories, capabilities building blocks and questions}
\label{tab:ccBbQuestions}
\endfirsthead
\endhead
 \hline
 \textbf{Capability Category} & \textbf{Capability Building Block (CBB)} & \textbf{Question} \\ 
 \hline
 \multirow{10}{8em}[-5em]{\textbf{Governance}} & Information Security Principles, Policies, and Controls & 1. To what extent are policies, laws, and guidelines for information security established? \\ 
 \cline{2-3}
  & Information Security Strategy & 2. To what extent are there guidelines and regulations for development and execution of information security strategies established? \\
 \cline{2-3}
  & Governance Structures & 3. To what extent are governance structures for  information security established? \\  
 \cline{2-3}
  & Roles, Responsibilities, and Accountabilities & 4. To what extent are responsibilities, penalties, incentive models, and audit options established? \\ 
 \cline{2-3}
  & Security Risk Management & 5. What approaches are in place for threat profiling and managing information security risks and vulnerabilities? \\ 
 \cline{2-3}
  & Skills and Competence Development & 6. To what extent is information security management training developed and disseminated? \\ 
 \cline{2-3}
  & Security Performance Measurement & 7. To what extent are key successful innovative programs, initiatives launched, and key factors and lessons learned for cybersecurity? \\ 
 \cline{2-3}
   & Cybersecurity Implementation Frameworks & 8. To what extent are implementation frameworks developed and standards adopted? \\ 
 \cline{2-3}
   & Cybersecurity Implementations Risks and Challenges & 9. To what extent are risks and challenges for cybersecurity  implementation plans identified and addressed? \\ 
 \cline{2-3}
   & Cybersecurity Technology Trends & 10. To what extent are the emerging technologies and future trends for cybersecurity identified? \\ 
 \hline
   \multirow{4}{8em}[-2em]{\textbf{Technical Security}} & Security Architecture & 11. How does the government define security architectures guidelines  and tools for cybersecurity? \\ 
 \cline{2-3}
   & IT Device Security & 12. To what extent are documents to define, implement, and monitor measures to protect IT devices established? \\ 
 \cline{2-3}
   & Cybersecurity Threats & 13. What information about threats is available at the country level? \\ 
 \cline{2-3}
   & Cybersecurity Threat Actors & 14. What information about threats actors is available at the country level? \\ 
 \cline{2-3}
   \multirow{4}{8em}[-3em]{\textbf{Security Data Administration}} & Data Security Classification & 15. To what extent are data security classification guidelines established? \\ 
  \hline
   & Data Life Cycle Management & 16. To what extent is data life cycle management addressed by the government? \\ 
 \cline{2-3}
   & Data Security Administration & 17. To what extent is data security administration addressed by the government? \\ 
 \cline{2-3}
   & Identity and Access Management & 18. To what extent are identity and data access management addressed by the government? \\ 
 \hline
   \multirow{2}{8em}[-0.3em]{\textbf{Business Continuity Management}} & Business Continuity Planning and Management & 19. What information security guidance is provided to support business continuity planning and management? \\ 
 \cline{2-3}
   & Incident Management & 20. To what extent is security incidents and near incidents management addressed by the government? \\ 
 \hline
\end{longtable}
}

\normalsize

\newpage

\subsection{Capability Maturity Levels}

\begin{figure}[!ht]
	\centering
	\includegraphics[width=1\linewidth]{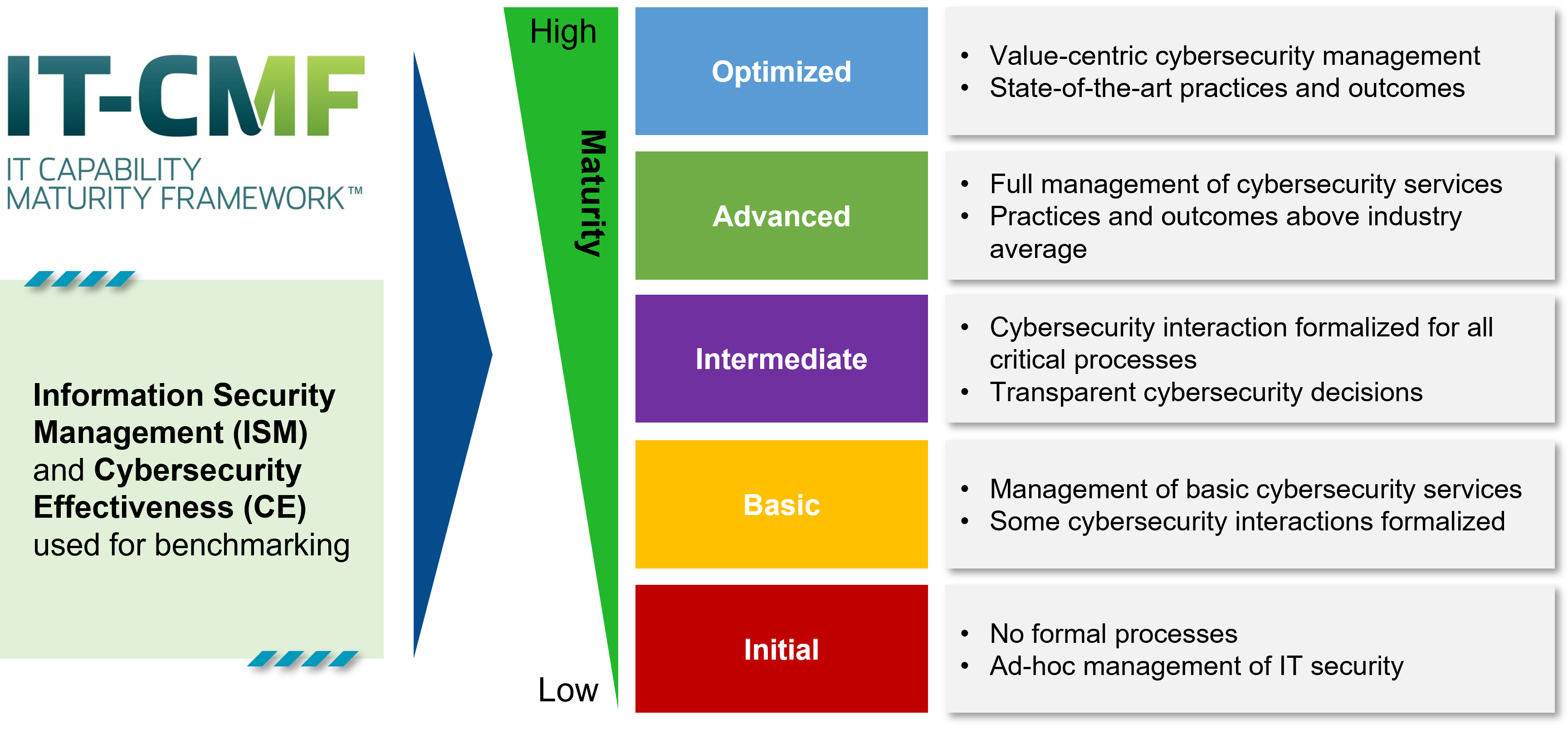}
	\caption{IT-CMF Capability Maturity Framework}
\end{figure}

The five maturity profiles for each question are reported in Table 2. Please notice that the five levels have the following meaning:

\begin{enumerate}
  \item Initial
  \item Basic
  \item Intermediate
   \item Advanced
    \item Optimized
\end{enumerate}

\footnotesize{
\begin{longtable}{ | m{4.5em} | m{13.5cm}|}  
\caption{Country-level maturity profiles}
\label{tab:ccBbLevels1}
\endfirsthead
\endhead
 \hline
{\cellcolor[RGB]{252,221,207}}\textbf{Maturity Question} & {\cellcolor[RGB]{252,221,207}}1. To what extent are policies, laws, and guidelines for information security established? \\
 \hline
 \textbf{Maturity Level} & \textbf{Description} \\
 \hline
 \multicolumn{1}{|c|}{1} & Information security policies, laws and guidelines are considered in an ad hoc manner, if at all (e.g., there is no national cybersecurity strategy). \\
 \hline
 \multicolumn{1}{|c|}{2} & Basic information security policies, laws, and guidelines are emerging (e.g., there is a basic national cybersecurity strategy). \\
 \hline
 \multicolumn{1}{|c|}{3}  & Formalised information security policies, laws, and guidelines are in place (e.g., there is a formalised national cybersecurity strategy). \\ 
\hline
 \multicolumn{1}{|c|}{4} & Comprehensive information security policies, laws, and guidelines are in place (e.g., there is a comprehensive national cybersecurity strategy). They are aligned with changes in strategic priorities, regulations, technology trends, and security risks. \\
 \hline
 \multicolumn{1}{|c|}{5} & Information security policies, laws, and guidelines are exemplary and continually updated (e.g., there is an exemplary and continually updated national cybersecurity strategy). They reflect the latest national security thinking and are reviewed for the national ecosystem, incorporating insights from the latest research from vendor recommendations,  technology trends, and lessons learned.  \\ 
 \hline
 
 {\cellcolor[RGB]{252,221,207}}\textbf{Maturity Question} & {\cellcolor[RGB]{252,221,207}}2. To what extent are there guidelines and regulations for development and execution of information security strategies established? \\
 \hline
 \textbf{Maturity Level} & \textbf{Description} \\
 \hline
 \multicolumn{1}{|c|}{1} & Providing ad hoc guidelines and regulations for development and execution of the national cybersecurity strategy, if at all (e.g., there is no implementation plan of the national cybersecurity strategy). \\
 \hline
 \multicolumn{1}{|c|}{2} & Basic  guidelines and regulations for development and execution of the national cybersecurity strategy are  emerging (e.g., there is a basic implementation plan of the national cybersecurity stratregy). \\
 \hline
 \multicolumn{1}{|c|}{3} & Formalised guidelines and regulations for development and execution of the  national cybersecurity strategy are in place (e.g., there is a formalised implementation plan of the national cybersecurity strategy). They are aligned with strategic priorities and security risks. \\ 
\hline
 \multicolumn{1}{|c|}{4} & Comprehensive guidelines and regulations for development and execution of the national cybersecurity strategy are in place (e.g., there is a comprehensive implementation plan of the national cybersecurity strategy).  They are aligned with changes in strategic priorities, security risk, regulations, standards, tools, and technologies. \\
 \hline
 \multicolumn{1}{|c|}{5} & Guidelines and regulations for development and execution of the national cybersecurity strategy are continually reviewed and updated (e.g., there is a continually updated implementation plan of the national cybersecurity strategy). They reflect the latest national security thinking and are aligned with changes in strategic priorities. They are reviewed for the national ecosystem, incorporating insights from the latest research from vendor recommendations, technology trends, and lessons learned.  \\ 
 \hline
 
 {\cellcolor[RGB]{252,221,207}}\textbf{Maturity Question} & {\cellcolor[RGB]{252,221,207}}3. To what extent are governance structures for  information security established? \\
 \hline
 \textbf{Maturity Level} & \textbf{Description} \\
 \hline
 \multicolumn{1}{|c|}{1} & Providing ad hoc governance, if at all (e.g., there is no governance structure for the coordination of the national cybersecurity). \\
 \hline
 \multicolumn{1}{|c|}{2} & Informal governance structures with participants mostly from a single government department (e.g., there is an informal governance structure for the coordination of the national cybersecurity). \\
 \hline
 \multicolumn{1}{|c|}{3} & Formal governance structures are in place (e.g., there is a formal governance structure for the coordination of the national cybersecurity).  There are participants from one or more government departments. \\ 
\hline
 \multicolumn{1}{|c|}{4} & Comprehensive governance structures are in place (e.g., there is a comprehensive governance structure with responsibilities for the coordination of the national cybersecurity). There is stakeholder participation with cross-functional collaboration. \\
 \hline
 \multicolumn{1}{|c|}{5} & Exemplary and continually updated governance structures are in place (e.g., there is an exemplary and continually updated governance structure with clear responsibilities for the coordination of the national cybersecurity). There is multiple-sector stakeholder participation and collaboration, informing the governance.  \\ 
  \hline
 
  {\cellcolor[RGB]{252,221,207}}\textbf{Maturity Question} & {\cellcolor[RGB]{252,221,207}}4. To what extent are responsibilities, penalties, incentive models, and audit options established? \\
 \hline
 \textbf{Maturity Level} & \textbf{Description} \\
 \hline
 \multicolumn{1}{|c|}{1} & Providing ad hoc information, if at all (e.g., there are no roles and responsibilities for government and operators of essential services/digital service providers). \\
 \hline
 \multicolumn{1}{|c|}{2} & Basic information/regulations emerging (e.g., basic roles and responsibilities for government and operators of essential services/digital service providers are defined). \\
 \hline
 \multicolumn{1}{|c|}{3} & Formalised information/regulations in place (e.g., formalised roles, responsibilities, and penalties for government and operators of essential services/digital service providers are established). They are aligned with strategic priorities and security management activities. \\ 
\hline
 \multicolumn{1}{|c|}{4} & Comprehensive information/regulations in place (e.g., comprehensive roles, responsibilities, and penalties for government and operators of essential services/digital service providers are established).  They are aligned with strategic priorities, security management activities, and standards. \\
 \hline
 \multicolumn{1}{|c|}{5} & Continually updated Information/initiatives (e.g., roles, responsibilities, and penalties for government and operators of essential services/digital service providers are fully established and continually revised/updated). They reflect the latest security thinking and strategic changes and they are informed by the national ecosystem.  \\ 
 \hline

 {\cellcolor[RGB]{252,221,207}}\textbf{Maturity Question} & {\cellcolor[RGB]{252,221,207}}5. What approaches are in place for threat profiling and managing information security risks and vulnerabilities? \\
 \hline
 \textbf{Maturity Level} & \textbf{Description} \\
 \hline
 \multicolumn{1}{|c|}{1} & Information security threat profiling and risk and vulnerability management approaches are considered in an ad hoc manner, if at all (e.g., there is no threats profiling information). \\
 \hline
 \multicolumn{1}{|c|}{2} & Some basic information security threat profiling and risk and vulnerability management are emerging (e.g., there is basic threats profiling information). \\
 \hline
 \multicolumn{1}{|c|}{3} & Formalised approaches for information security threat profiling and risk and vulnerability management are established (e.g., there is formalised threats profiling information). Improvements are also made based on changes in the risk landscape. \\ 
\hline
 \multicolumn{1}{|c|}{4} & The information security threat profiling and risk and vulnerability management approaches are comprehensive  (e.g., there is  comprehensive threats profiling information). Improvements are also made based on changes in the risk landscape and from lessons learned from previous information security incidents. \\
 \hline
 \multicolumn{1}{|c|}{5} & The information security threat profiling  and risk and vulnerability management approaches are continually informed by the latest recommendations from security agencies, vendors, and emerging research concepts (e.g., there is continually updated threats profiling information). Improvements are also made based on changes in the risk landscape and from lessons learned from previous information security incidents across the national ecosystem.  \\ 
 \hline
   
 {\cellcolor[RGB]{252,221,207}}\textbf{Maturity Question} & {\cellcolor[RGB]{252,221,207}}6. To what extent is information security management training developed and disseminated? \\
 \hline
 \textbf{Maturity Level} & \textbf{Description} \\
 \hline
 \multicolumn{1}{|c|}{1} & Providing few, if any, training programmes  (e.g., there are no cybersecurity training programmes provided by the government). \\
 \hline
 \multicolumn{1}{|c|}{2} & Some basic training programmes established mainly focusing on cybersecurity awareness (e.g., basic cybersecurity training programmes are provided by the government). \\
 \hline
 \multicolumn{1}{|c|}{3} & Formal training programmes are in place and they are mainly focusing on strategic priorities (e.g., formal cybersecurity training and certification programmes are provided by the government based on security issues and threats).  \\ 
\hline
 \multicolumn{1}{|c|}{4} & Comprehensive cybersecurity training and development programmes are based on strategic priorities and knowledge gaps (e.g., comprehensive cybersecurity training and certification programmes are provided by the government for different audiences). They reflect the latest tools, technologies, and methods. \\
 \hline
 \multicolumn{1}{|c|}{5} & Cybersecurity training and development programmes are continually reviewed and improved based on strategic priorities and are adjusted as required to address knowledge gaps (e.g., full cybersecurity training and certification programmes are provided by the government for all different audiences and continually updated based on the national ecosystem). They reflect the latest tools, technologies, methods, and emerging research concepts in cybersecurity management.  \\ 
 \hline

 {\cellcolor[RGB]{252,221,207}}\textbf{Maturity Question} & {\cellcolor[RGB]{252,221,207}}7. To what extent are key successful innovative programs, initiatives launched, and key factors and lessons learned for cybersecurity? \\
 \hline
 \textbf{Maturity Level} & \textbf{Description} \\
 \hline
 \multicolumn{1}{|c|}{1} & Information security innovative programs and initiatives are considered in an ad hoc manner, if at all (e.g., there is no report on cybersecurity lessons learned). \\
 \hline
 \multicolumn{1}{|c|}{2} & Basic information security innovative programs and initiatives are emerging. They incorporate some insights from the lessons learned  (e.g., there is a basic report on cybersecurity  lessons learned). \\
 \hline
 \multicolumn{1}{|c|}{3} & Formalised information security innovative programs and initiatives in place. They incorporate insights from the lessons learned  (e.g., there is a formalised report on cybersecurity  lessons learned).  \\ 
\hline
 \multicolumn{1}{|c|}{4} & Comprehensive information security innovative programs and initiatives are in place. They reflect the latest national security thinking and incorporate insights from the lessons learned  (e.g., there is a comprehensive report on cybersecurity lessons learned from multiple sectors). \\
 \hline
 \multicolumn{1}{|c|}{5} & Information security innovative programs and initiatives are continually reviewed. They reflect the latest national security thinking and are reviewed for the national ecosystem,  incorporating insights from the latest research from vendor recommendations, technology trends, and lessons learned  (e.g., there is an exemplary and continually updated report on cybersecurity lessons learned from all sectors).  \\ 
 \hline 
  
 {\cellcolor[RGB]{252,221,207}}\textbf{Maturity Question} & {\cellcolor[RGB]{252,221,207}}8. To what extent are implementation frameworks developed and standards adopted? \\
 \hline
 \textbf{Maturity Level} & \textbf{Description} \\
 \hline
 \multicolumn{1}{|c|}{1} & Providing ad hoc information, if at all (e.g., there is no  cybersecurity implementation framework for organisations recommeded by the government). \\
 \hline
 \multicolumn{1}{|c|}{2} & Basic implementation framework is emerging (e.g., there is a basic cybersecurity implementation framework for organisations recommended by the government). \\
 \hline
 \multicolumn{1}{|c|}{3} & Formalised implementation framework in place (e.g., there is a formalised cybersecurity implementation  framework for organisations recommended by the government).  Its is aligned with security standards.  \\ 
\hline
 \multicolumn{1}{|c|}{4} & Comprehensive implementation framework in place (e.g., there is a comprehensive cybersecurity implementation framework for organisations recommended by the governement). It is aligned with the latest standards and reviewed for the national practices, incorporating insights from vendor recommendations and technology trends. \\
 \hline
 \multicolumn{1}{|c|}{5} & The implementation framework is continually reviewed and updated (e.g., there is a continually updated cybersecurity framework for organisations recommended by the goverment). It is fully aligned with the latest standards and reviewed for the national ecosystem and practices,  incorporating insights from the latest research from vendor recommendations, technology trends, and lessons learned.  \\ 
   \hline 
   
  {\cellcolor[RGB]{252,221,207}}\textbf{Maturity Question} & {\cellcolor[RGB]{252,221,207}}9. To what extent are risks and challenges for cybersecurity implementation plans identified and addressed? \\
 \hline
 \textbf{Maturity Level} & \textbf{Description} \\
 \hline
 \multicolumn{1}{|c|}{1} & Risks and challenges are considered in an ad hoc manner, if at all (e.g., there is no report on cybersecurity challenges and risks). \\
 \hline
 \multicolumn{1}{|c|}{2} & Basic risks and challenges for cybersecurity implementation plans identified by the government are in place (e.g., there is a basic report on cybersecurity challenges and risks). \\
 \hline
 \multicolumn{1}{|c|}{3} & 	Formalised risks and challenges for cybersecurity implementation plans identified by the government are in place (e.g., there is a formalised report on cybersecurity challenges and risks). They are based on some cybersecurity priority problems of the national ecosystem.  \\ 
\hline
 \multicolumn{1}{|c|}{4} & Comprehensive risks and challenges for cybersecurity implementation plans identified by the government are in place, using information from security agencies and vendors (e.g., there is a comprehnsive report on cybersecurity challenges, associated risks, and mitigations). They are based on the highest cybersecurity priority problems of the national ecosystem.  \\
 \hline
 \multicolumn{1}{|c|}{5} & Risks and challenges for cybersecurity implementation plans are continually informed and addressed by the government, using the latest information from security agencies, vendors, and emerging research concepts (e.g., there is a continually updated report on cybersecurity challenges, associated risks, and mitigations). They are based on the highest cybersecurity priority problems and from lessons learned from previous information security incidents and threats across the national ecosystem.  \\ 
   \hline 
  
 {\cellcolor[RGB]{252,221,207}}\textbf{Maturity Question} & {\cellcolor[RGB]{252,221,207}}10. To what extent are the emerging technologies and future trends for cybersecurity identified? \\
 \hline
 \textbf{Maturity Level} & \textbf{Description} \\
 \hline
 \multicolumn{1}{|c|}{1} & Emerging technologies and future trends are considered in an ad hoc manner, if at all (e.g., there is no report on emerging technologies and future trends). \\
 \hline
 \multicolumn{1}{|c|}{2} & Emerging technologies and future trends are basically identified by the government (e.g., there is a basic report on emerging technologies and future trends). \\
 \hline
 \multicolumn{1}{|c|}{3} & Emerging technologies and future trends are formally identified by the government (e.g., there is a formalised report on emerging technologies and future trends). They reflect some insights and innovations affecting cybersecurity.  \\ 
\hline
 \multicolumn{1}{|c|}{4} & Emerging technologies and future trends are comprehensively identified by the government, using the latest information from security agencies, and vendors (e.g., there is a comprehensive report on emerging technologies and future trends,   associated vulnerabilities and risks). They reflect strategic priorities and insights and innovations affecting cybersecurity in multiple sectors.  \\
 \hline
 \multicolumn{1}{|c|}{5} & Emerging technologies and future trends are fully identified by the government, using the latest information from security agencies, and vendors (e.g., there is a continually updated report on emerging technologies and future trends, associated vulnerabilities, and risks). They fully reflect strategic priorities and insights and innovations affecting cybersecurity across the whole national ecosystem.  \\ 
   \hline 
  
 {\cellcolor[RGB]{252,221,207}}\textbf{Maturity Question} & {\cellcolor[RGB]{252,221,207}}11. How does the government define security architectures guidelines and tools for cybersecurity? \\
 \hline
 \textbf{Maturity Level} & \textbf{Description} \\
 \hline
  \multicolumn{1}{|c|}{1} & Providing ad hoc security architecture guidelines and tools, if at all (e.g., there is no cybersecurity architecture document). \\
 \hline
 \multicolumn{1}{|c|}{2} & Basic security architecture guidelines and tools are in place (e.g., there is a basic cybersecurity architecture document).  \\
 \hline
 \multicolumn{1}{|c|}{3} & Formalised security architecture guidelines and tools are in place (e.g., there is a formalised cybersecurity architecture document). They are aligned with standards.  \\ 
\hline
 \multicolumn{1}{|c|}{4} & Comprehensive security architecture guidelines and tools are in place (e.g., there is a comprehensive cybersecurity architecture document). They are aligned with standards and incorporate insights from vendor recommendations and technology trends.  \\
 \hline
 \multicolumn{1}{|c|}{5} & The security architecture guidelines and tools are continually reviewed and updated (e.g., there is a continually updated cybersecurity architecture document). They are fully aligned with the latest standards and reviewed for the national ecosystem and practices,  incorporating insights from the latest research from vendor recommendations, technology trends, and lessons learned.  \\ 
   \hline 
 
  {\cellcolor[RGB]{252,221,207}}\textbf{Maturity Question} & {\cellcolor[RGB]{252,221,207}}12. To what extent are documents to define, implement, and monitor measures to protect IT devices established? \\
 \hline
 \textbf{Maturity Level} & \textbf{Description} \\
 \hline
  \multicolumn{1}{|c|}{1} & Providing ad hoc definition, implementation and monitoring measures to protect IT devices, if at all (e.g., there is no guideline with security measures to protect IT devices). \\
 \hline
 \multicolumn{1}{|c|}{2} & Basic definition, implementation and monitoring measures to protect IT devices are in place (e.g., there is a basic guideline with security measures to protect IT devices).  \\
 \hline
 \multicolumn{1}{|c|}{3} & Formalised definition, implementation and monitoring measures to protect IT devices are in place (e.g., there is a formalised guideline with security measures to protect some IT devices). They are aligned with standards.  \\ 
\hline
 \multicolumn{1}{|c|}{4} & Comprehensive definition, implementation and monitoring measures to protect IT devices are in place (e.g., there is a comprehensive guideline with security measures to protect most IT devices). They are aligned with standards and incorporate insights from vendor recommendations and technology trends.  \\
 \hline
 \multicolumn{1}{|c|}{5} & Definition, implementation and monitoring measures to protect all IT devices are continually reviewed and updated (e.g., there is a continually updated guideline with security measures to protect all IT devices). They are fully aligned with the latest standards and reviewed for the national ecosystem and practices, incorporating insights from the latest research from vendor recommendations, technology trends, and lessons learned.  \\ 
   \hline
 
  {\cellcolor[RGB]{252,221,207}}\textbf{Maturity Question} & {\cellcolor[RGB]{252,221,207}}13. What information about threats is available at the country level? \\
 \hline
 \textbf{Maturity Level} & \textbf{Description} \\
 \hline
 \multicolumn{1}{|c|}{1} & Providing ad hoc information about cybersecurity threats, if at all (e.g., there is no report on cybersecurity threats). \\
 \hline
 \multicolumn{1}{|c|}{2} & Basic information about cybersecurity threats is in place (e.g., there is a basic report on cybersecurity threats).  \\
 \hline
 \multicolumn{1}{|c|}{3} & Formalised information about cybersecurity threats is in place (e.g., there is a formalised report on cybersecurity threats and descriptive information). It reflects the national security thinking.  \\ 
\hline
 \multicolumn{1}{|c|}{4} & Comprehensive information about cybersecurity threats is in place, provided for different audiencies, and informed by the feedback from security agencies and vendors (e.g., there is a comprehensive report on cybersecurity threats, associated risks, and advice). It reflects the latest national security thinking.  \\
 \hline
 \multicolumn{1}{|c|}{5} & Full information about cybersecurity threats is continually updated, provided for all different audiencies, and  informed by the latest feedback from security agencies, vendors, and emerging research (e.g., there is a full and continually updated report on cybersecurity threats, associated risks, and solutions). It fully reflects the latest national security thinking and is reviewed for the national ecosystem.  \\ 
 \hline
 
 {\cellcolor[RGB]{252,221,207}}\textbf{Maturity Question} & {\cellcolor[RGB]{252,221,207}}14. What information about threats actors is available at the country level? \\
 \hline
 \textbf{Maturity Level} & \textbf{Description} \\
 \hline
 \multicolumn{1}{|c|}{1} & Providing ad hoc information about cybersecurity threat actors, if at all (e.g., there is no report on cybersecurity threat actors). \\
 \hline
 \multicolumn{1}{|c|}{2} & Basic information about cybersecurity threat actors is in place (e.g., there is a basic report on cybersecurity threat actors).  \\
 \hline
 \multicolumn{1}{|c|}{3} & Formalised information about cybersecurity threat actors is in place (e.g., there is a formalised report on cybersecurity threat actors and descriptive information). It reflects the national security thinking.  \\ 
\hline
 \multicolumn{1}{|c|}{4} & Comprehensive information about cybersecurity threat actors is in place, provided for different audiencies, and informed by the feedback from security agencies and vendors (e.g., there is a comprehensive report on cybersecurity threat actors, associated risks, and advice). It reflects the latest national security thinking.  \\
 \hline
 \multicolumn{1}{|c|}{5} & Full information about cybersecurity threat actors is continually updated, provided for all different audiencies, and  informed by the latest feedback from security agencies, vendors, and emerging research (e.g., there is a full and continually updated report on cybersecurity threat actors, associated risks, and solutions). It fully reflects the latest national security thinking and is reviewed for the national ecosystem.  \\ 
 \hline
 
 {\cellcolor[RGB]{252,221,207}}\textbf{Maturity Question} & {\cellcolor[RGB]{252,221,207}}15. To what extent are data security classification guidelines established? \\
 \hline
 \textbf{Maturity Level} & \textbf{Description} \\
 \hline
 \multicolumn{1}{|c|}{1} & Providing ad hoc data security classification guidelines, if at all (e.g., there is no guideline/manual for data classification). \\
 \hline
 \multicolumn{1}{|c|}{2} & Basic data security classification guidelines are in place (e.g., there is a basic guideline/manual for data classification).  \\
 \hline
 \multicolumn{1}{|c|}{3} & Formalised  data security classification guidelines are in place (e.g., there is a formalised guideline/manual for data classification). They are aligned with data standards.  \\ 
\hline
 \multicolumn{1}{|c|}{4} & Comprehensive data security classification guidelines are in place (e.g., there is a comprehensive guideline/manual for data classification). They are aligned with data standards and incorporate insights from vendor recommendations and technology trends.  \\
 \hline
 \multicolumn{1}{|c|}{5} & Data security classification guidelines  are continually reviewed and updated (e.g., there is a continually updated guideline/manual for data classification). They are fully aligned with the latest data standards and reviewed for the national ecosystem and practices, incorporating insights from the latest research from vendor recommendations, technology trends, and lessons learned.  \\ 
 \hline

 {\cellcolor[RGB]{252,221,207}}\textbf{Maturity Question} & {\cellcolor[RGB]{252,221,207}}16. To what extent is data life cycle management addressed by the government? \\
 \hline
 \textbf{Maturity Level} & \textbf{Description} \\
 \hline
 \multicolumn{1}{|c|}{1} & Providing ad hoc documents for managing the security of data life cycle, if at all (e.g., there is no document to guide the management of data life cycle). \\
 \hline
 \multicolumn{1}{|c|}{2} & Basic documents for managing the security of data life cycle are in place (e.g., there is a basic document to guide the management of data life cycle).  \\
 \hline
 \multicolumn{1}{|c|}{3} & Formalised  data security classification guidelines are in place (e.g., there is a formalised guideline/manual for data classification). They are aligned with data standards.  \\ 
\hline
 \multicolumn{1}{|c|}{4} & Comprehensive documents for managing the security of data life cycle are in place (e.g., there is a comprehensive document to guide the management of each stage of the data life cycle). They are aligned with data standards and incorporate insights from vendor recommendations and technology trends.  \\
 \hline
 \multicolumn{1}{|c|}{5} & 	Documents for managing the security of data life cycle are continually reviewed and updated (e.g., there is a continually updated document to guide the management of each stage of the data life cycle). They are fully aligned with the latest data standards and reviewed for the national ecosystem and practices, incorporating insights from the latest research from vendor recommendations, technology trends, and lessons learned.  \\ 
 \hline

 {\cellcolor[RGB]{252,221,207}}\textbf{Maturity Question} & {\cellcolor[RGB]{252,221,207}}17. To what extent is data security administration addressed by the government? \\
 \hline
 \textbf{Maturity Level} & \textbf{Description} \\
 \hline
 \multicolumn{1}{|c|}{1} & Providing ad hoc data security adminstration regulations/guidelines, if at all (e.g., there is no goverment regulation for personal data protection). \\
 \hline
 \multicolumn{1}{|c|}{2} & Basic data security administration regulations/guidelines are in place (e.g., there is a basic goverment regulation for personal data protection).  \\
 \hline
 \multicolumn{1}{|c|}{3} & Formalised  data security adminis tration regulations/guidelines are in place (e.g., there is a formalised goverment regulation for personal data protection). They are aligned with and standards.  \\ 
\hline
 \multicolumn{1}{|c|}{4} & Comprehensive data security administration regulations/guidelines are in place (e.g., there is a comprehensive goverment regulation for personal data protection). They are aligned with standards and incorporate insights from vendor recommendations and technology trends.  \\
 \hline
 \multicolumn{1}{|c|}{5} & 	Data security administration regulations/guidelines are continually reviewed and updated (e.g., there is a continually updated goverment regulation for personal data protection). They are fully aligned with the latest standards and reviewed for the national ecosystem and practices, incorporating insights from the latest research from vendor recommendations, technology trends, and lessons learned.  \\ 
 \hline

 {\cellcolor[RGB]{252,221,207}}\textbf{Maturity Question} & {\cellcolor[RGB]{252,221,207}}18. To what extent are identity and data access management
addressed by the government? \\
 \hline
 \textbf{Maturity Level} & \textbf{Description} \\
 \hline
 \multicolumn{1}{|c|}{1} & Providing ad hoc identity and data access management frameworks, guidelines, and systems, if at all (e.g., there is no guideline for e-identity). \\
 \hline
 \multicolumn{1}{|c|}{2} & Basic identity and data access management frameworks, guidelines, and systems are in place (e.g., there is a basic guideline for e-identity).  \\
 \hline
 \multicolumn{1}{|c|}{3} & 	Formalised identity and data access management frameworks, guidelines, and systems are in place (e.g., there is a formalised guideline/system for e-identity). They are aligned with regulations and standards.  \\ 
\hline
 \multicolumn{1}{|c|}{4} & Comprehensive identity and data access management frameworks, guidelines, and systems are in place (e.g., there is a comprehensive guideline and system for e-identity). They are aligned with regulations, standards and incorporate insights from vendor recommendations and technology trends.  \\
 \hline
 \multicolumn{1}{|c|}{5} & 	Identity and data access management frameworks, guidelines, and systems are continually reviewed and updated (e.g., there is a continually updated guidelines and systems for e-identity). They are fully aligned with the latest regulations, standards and reviewed for the national ecosystem and practices, incorporating insights from the latest research from vendor recommendations, technology trends, and lessons learned.  \\ 
 \hline

 {\cellcolor[RGB]{252,221,207}}\textbf{Maturity Question} & {\cellcolor[RGB]{252,221,207}} 19. What information security guidance is provided to support business continuity planning and management? \\
 \hline
 \textbf{Maturity Level} & \textbf{Description} \\
 \hline
 \multicolumn{1}{|c|}{1} & Information security guidance to support business continuity is considered in an ad hoc manner, if at all (e.g., there is no management plan for business continuity). \\
 \hline
 \multicolumn{1}{|c|}{2} & 	Basic information security guidance is provided to support business continuity. There is a basic national contingency and recovery plan (e.g., there is a basic management plan for business continuity).  \\
 \hline
 \multicolumn{1}{|c|}{3} & 	Formalised identity and data access management frameworks, guidelines, and systems are in place (e.g., there is a formalised guideline/system for e-identity). They are aligned with regulations and standards.  \\ 
\hline
 \multicolumn{1}{|c|}{4} &	Formalised information security guidance is provided to support business continuity. There is a national cyber crisis unit responsible for the implementation of contingency and recovery plan (e.g., there is a formalised  cyber crisis unit and management plan for business continuity).  \\
 \hline
 \multicolumn{1}{|c|}{5} & 	Comprehensive information security guidance is provided to support business continuity. There is a national cyber crisis unit responsible for the execution of cyber crisis exercises and the implementation of contingency and recovery plan, informed by regulations (e.g., there is a comprehensive  cyber crisis unit and management plan for business continuity).	Information security guidance is provided to support business continuity and it is continually and proactively reviewed and improved. There is a continually reviewed national cyber crisis unit responsible for the execution of collective cyber crisis exercises and the implementation of contingency and recovery plan, informed by regulations (e.g., there is a continually reviewed and updated  cyber crisis unit and management plan for business continuity).  \\ 
 \hline

 {\cellcolor[RGB]{252,221,207}}\textbf{Maturity Question} & {\cellcolor[RGB]{252,221,207}}20. To what extent is security incidents and near incidents management addressed by the government? \\
 \hline
 \textbf{Maturity Level} & \textbf{Description} \\
 \hline
 \multicolumn{1}{|c|}{1} & Incident management approaches are considered in an ad hoc manner, if at all (e.g., there is no incident response centre at the national level). \\
 \hline
 \multicolumn{1}{|c|}{2} & 	Basic approaches emerging for managing a limited number of incident types (e.g., there is a basic incident response centre that receives limited incident notifications and manages the related incidents).  \\
 \hline
 \multicolumn{1}{|c|}{3} & 	Formalised approaches used for managing most IT security incidents (e.g., there is a formalised incident response centre that receives incident notifications and manages incidents). Most incidents are systematically managed.  \\ 
\hline
 \multicolumn{1}{|c|}{4} &	Advanced approaches are used for all IT security incidents (e.g., there is an advanced incident response centre that is coordinated with competent government bodies, receives incident notifications from different sectors/companies, and manages incidents).  Incidents are thoroughly and systematically managed.  \\
 \hline
 \multicolumn{1}{|c|}{5} & 	Full approaches to incident management are continually and proactively reviewed and improved. They are informed by research, by recommendations from security agencies, vendors, and lessons learned  (e.g., there is a continually informed incident response centre that is coordinated with all competent government bodies, receives incident notifications from all sectors/companies, and manages incidents by using latest research and recommendations).  \\ 
 \hline

\end{longtable}
}

\subsection{Qualitative Questions}

Table~\ref{tab:QualitativeQuestions} presents the list of questions by using a qualitative perspective of cybersecurity aspects which should be assessed at the national level.

\footnotesize
\begin{longtable}{ | m{1.5em}| m{13.5cm} |} 
\caption{Qualitative questions}
\label{tab:QualitativeQuestions}
\endfirsthead
\endhead
\hline
 \textbf{No.} & \textbf{Qualitative Questions}  \\ 
\hline
 
\multicolumn{1}{|r|}{1.} & What is the government cybersecurity policy unit?\\ 
 \hline
 
\multicolumn{1}{|r|}{2.} &	What are the government committees, working groups, etc. for cybersecurity?\\ 
 \hline
	
\multicolumn{1}{|r|}{3.} &	What is the government national-level cybersecurity strategy or other equivalent document?\\ 
 \hline
	
\multicolumn{1}{|r|}{4.} &	What is the government implementation plan to the national-level cybersecurity strategy or other equivalent document?\\ 
 \hline
	
\multicolumn{1}{|r|}{5.} &	What is the government entity unit that is specialised in national strategic cyber threats?\\ 
 \hline

\multicolumn{1}{|r|}{6.} &	What are the cyber threat reports published by the government?\\ 
 \hline
	
\multicolumn{1}{|r|}{7.} &	What are the government cyber safety and security websites?\\
 \hline
	
\multicolumn{1}{|r|}{8.} &	What is the training on cibersecurity offered by government?\\ 
 \hline
	
\multicolumn{1}{|r|}{9.} &	What are the international cooperation activites of government dedicated to cybersecurity  (e.g. FIRST)?\\ 
 \hline
	
\multicolumn{1}{|r|}{10.} &	What are the government cybersecurity guidelines/notes for the digital service providers?\\ 
 \hline
	
\multicolumn{1}{|r|}{11.} &	What are the government cybersecurity guidelines/notes for operators of essential services?\\ 
 \hline
	
\multicolumn{1}{|r|}{12.} &	What evidence of the effective implementation of cyber/information security policies (e.g. audit result, documentation, specific report) is provided by operators of essential services?\\ 
 \hline
	
\multicolumn{1}{|r|}{13.} &	What are the government legal acts for personal data protection?\\ 
 \hline
	
\multicolumn{1}{|r|}{14.} &	What are the government units (CSIRT, CERT, CIRT, etc.) specialised in national-level cyber incident detection and response?\\ 
 \hline

\multicolumn{1}{|r|}{15.} &	What are the government authorities that digital service providers and operators of essential services must notify of cybersecurity incidents?\\ 
 \hline

\multicolumn{1}{|r|}{16.} &	What is the government crisis management plan for large-scale cyber incidents?\\ 
 \hline
	
\multicolumn{1}{|r|}{17.} &	What are the cybersecurity standards referenced by the government?\\ 
 \hline
	
\multicolumn{1}{|r|}{18.} &	Is there any specific cybersecurity information for emerging technologies (AI, ML, etc.), cloud services, and operational technologies (OT) (e.g., asset management, OT devices)?\\ 
 \hline

\end{longtable}

%% file: irelandAssessment.tex
\section{National Cybersecurity Assessment: The Case of Ireland}

\subsection{Overall National Results}
This Section presents the results obtained by applying the methodology and instrument presented above to evaluate the cybersecurity maturity level of Ireland.

\begin{figure}[!ht]
	\centering
	\includegraphics[width=1.1\linewidth]{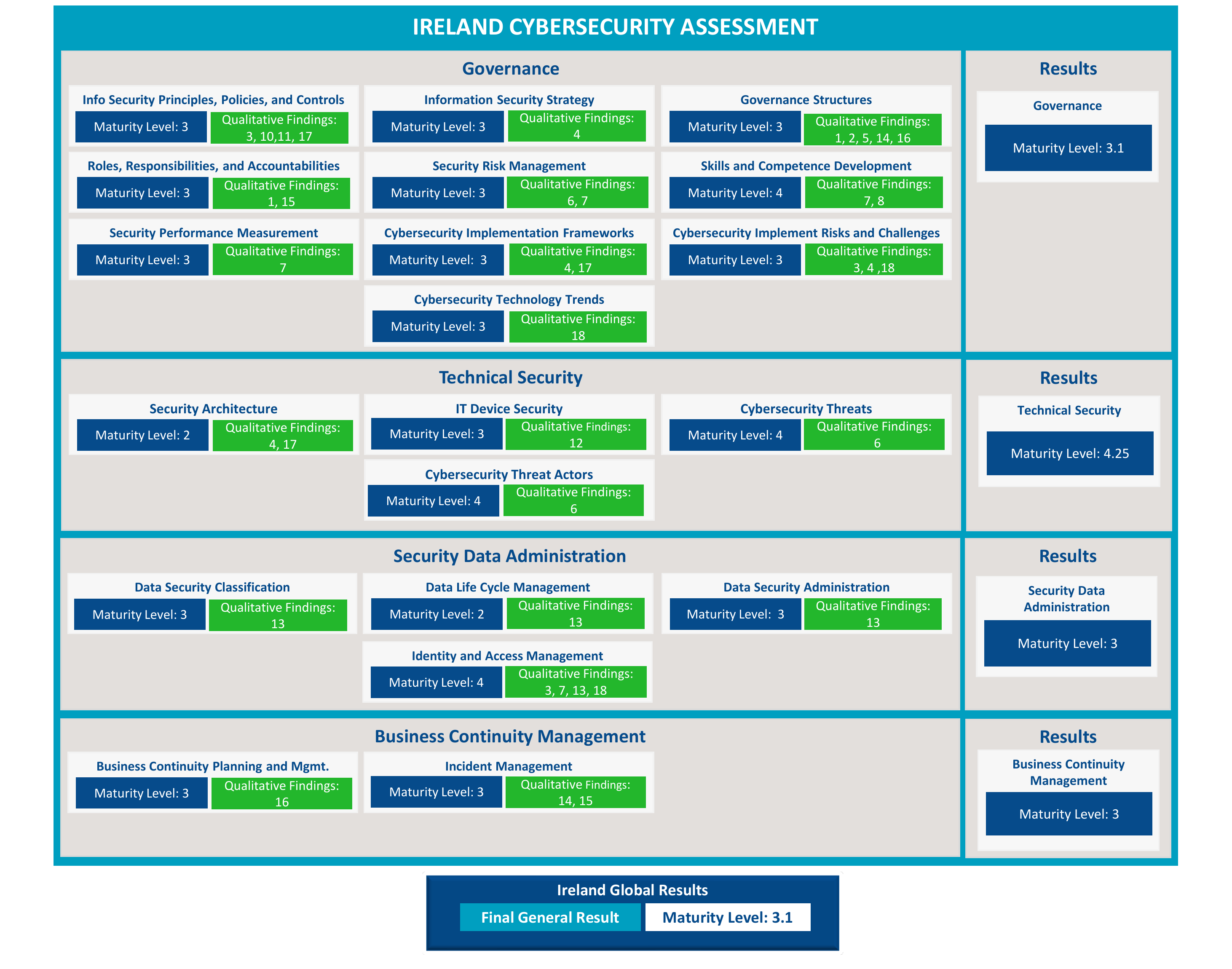}
	\caption{Assessment Results - Ireland}
	\label{fig:results}
\end{figure}

Figure~\ref{fig:results} reports the maturity scores in the four capability categories, Governance, Technical Security, Security Data Administration and Business Continuity Management, and the overall Ireland maturity score.

\subsection{Documents}

Table~\ref{tab:documents} presents the documents related to cybersecurity at the national level in Ireland used to evaluate the cybersecurity capabilities in this country.

\footnotesize{
\begin{longtable}{ | m{1em} | m{2cm}| m{3cm} | m{0.6cm} | m{6.6cm} |} 
\caption{Documents}
\label{tab:documents}
\endfirsthead

\hline
{\cellcolor[RGB]{0,152,187}}\color{white}{\textbf{No}} & {\cellcolor[RGB]{0,152,187}}\color{white}{\textbf{Short Name}} & {\cellcolor[RGB]{0,152,187}}\color{white}{\textbf{Long Name}} & {\cellcolor[RGB]{0,152,187}}\color{white}{\textbf{Year}} & {\cellcolor[RGB]{0,152,187}}\color{white}{\textbf{Description}}
\endhead

\hline
{\cellcolor[RGB]{0,152,187}}\color{white}\textbf{No} & {\cellcolor[RGB]{0,152,187}}\color{white}\textbf{Short Name} & {\cellcolor[RGB]{0,152,187}}\color{white}\textbf{Long Name} & {\cellcolor[RGB]{0,152,187}}\color{white}\textbf{Year} & {\cellcolor[RGB]{0,152,187}}\color{white}\textbf{Description}

\\ \hline

1  & AmCham GCS Report  & American Chamber Global CyberSecurity Leadership  & 2019 & Position paper of the American   Chamber with recommendations ofr Ireland in adopting a policy approach to   cyber security that is risk-based; outcome focused; respectful of civil liberties; internationally relevant;   adopts international best-practice and embeds      security-by-design.                                                                                                                                                                                                                                                                                                                                                                                                                                                                                                                                                                                                                                                                                                                                                                                                         \\ \hline
2  & CS Asset Management Report                                             & Thematic Inspection of   Cybersecurity Risk Management in Asset Management Firms & 2020 & Thematic Inspection of   Cybersecurity Risk Management in Asset Management Firms                    

\\ \hline
3  & IS Management                                                          & Advice Note: Considering   Information Security Management                                                                                                                                              & 2018 & The Advice Note: Considering   Information Security Management Supporting the Public Service ICT Strategy   contains the suggested strategies and activities to enhance information   security. t aims to assist Public Bodies in making informed, risk-based   decisions in relation to the improvement of Information Security within their   organisation.        
\\ \hline
4  & CS Guidance for Business                                               & Guidance on Cybersecurity for   Irish Business                                                                                                                                                          & 2018 & The 12 Steps to Cyber Security   Guidance on Cyber Security for Irish Business is intended to be used by   businesses as a suggested activity plan which may be undertaken on a   month-by-month basis over a suggested 12 month period to improve cyber   resilience.                                                                                                                                                                                                                                                                                                                                                                                                                                                                                                                                                                                                                                                                                                                                                                                                                                                                                        \\ \hline
5  & CS Legal Guide                                                         & International Comparative Legal   Guide (ICLG) to Cybersecurity                                                                                                                                         & 2018 & ICLG Cybersecurity covers common   issues in cybersecurity laws and regulations, including cybercrime,   applicable laws, preventing attacks, specific sectors, corporate governance,   litigation, insurance, and investigatory and police powers.                                                                                                                                                                                                                                                                                                                                                                                                                                                                                                                                                                                                                                                                                                                                                                                                                                                                                                           \\ \hline
6  & Incident Report Form                                                   & Digital Service Providers   Incident Notification Form                                                                                                                                                  & 2021 & Incident Notification Form to be   used by Digital Service Providers                                                                                                                                                                                                                                                                                                                                                                                                                                                                                                                                                                                                                                                                                                                                                                                                                                                                                                                                                                                                                                                                                          \\ \hline
7  & Digital Service Providers                                              & Information Note for Digital   Service Providers                                                                                                                                                        & 2018 & The Information Note for Digital   Service Providers assists Digital Service Providers in understanding their   obligations in relation to compliance with the NIS Directive.                                                                                                                                                                                                                                                                                                                                                                                                                                                                                                                                                                                                                                                                                                                                                                                                                                                                                                                                                                                 \\ \hline
8  & National CS Strategy                                                   & National Cyber Security Strategy                                                                                                                                                                        & 2019 & The National Cyber Security   Strategy was published in December 2019, and follows on from the country's   first strategy. It is a broader and more comprehensive document than the last   one, and is informed by the operational experience gained by the National   Cyber Security Centre (NCSC) from 2015 to 2019, and from ongoing national and   international engagements in the area. The vision behind the 2019 Strategy is for Ireland to continue to safely   enjoy the benefits of the digital revolution and to play a full part in   shaping the future of the Internet. This vision will be achieved   through: - the protection of the State, its people, and its critical national   infrastructure from cyber threats - the development of the capacity of the State, of research institutions,   of businesses and of the people, to both better understand and manage the   nature of the challenges we face - the engagement by the State, nationally and internationally, in a   strategic manner, supporting a free, open, peaceful and secure cyber space \\ \hline
9  & Operators of Essential Service                                         & NIS Compliance Guidelines for   Operators of Essential Service (OES)                                                                                                                                    & 2019 & The NIS Compliance Guidelines   for Operators of Essential Service (OES) establishes a set of Guidelines   designed to assist OES in meeting their network and information system   security and incident reporting requirements under Regulations 17 and 18 of   the NIS Regulations.                                                                                                                                                                                                                                                                                                                                                                                                                                                                                                                                                                                                                                                                                                                                                                                                                                                                        \\ \hline
10 & Requirements for CSIRTs                                                & Mandatory Requirements for   CSIRTs                                                                                                                                                                     & 2020 & It contains a description of the   National Cyber Security Centre of Ireland's (NCSC-IE) Computer Security   Incident Response Team (CSIRT-IE) in accordance with RFC 2350. It provides   basic information about the CSIRT-IE team, its channels of communication and   its roles and responsibilities.                                                                                                                                                                     \\ \hline
11 & Incident notification for DSPs                                         & Incident notification for DSPs   in the context of the NIS Directive                                                                                                                                    & 2017 & ENISA guideline on how to   implement incident notification for Digi-tal Service Providers, in the   context of the NIS Directive                                                                                                                                                                                                                                                                                                                                                                                                                                                                                                                                                                                                                                                                                                                                                                                                                                                                                                                                                                                                                             \\ \hline
12 & Data Protection Regulation                                             & Data Protection Regulation                                                                                                                                                                              &      &                                                                                                                                                                                                                                                                                                                                                                                                                                                                                                                                                                                                                                                                                                                                                                                                                                                                                                                                                                                                                                                                                                                                                               \\ \hline
13 & Security Regulation and Law                                            & S.I. No. 360/2018 - European   Union (Measures for a High Common Level of Security of Network and   Information Systems) Regulations 2018                                                               & 2018 & Statutory Instrument No. 360 of   2018 EUROPEAN UNION (MEASURES FOR A HIGH COMMON LEVEL OF SECURITY OF NETWORK   AND INFORMATION SYSTEMS)      REGULATIONS 2018                                                                                                                                                                                                                                                                                                                                                                                                                                                                                                                                                                                                                                                                                                                                                                                                                                                                                                                                                  \\ \hline
14 & NCS Measures Report                                                    & Report on the Accounts of the   Public Services 2017 (Chapter 8)                                                                                                                                        & 2017 & Report on the Accounts of the   Public Services 2017 containing the measures relating to national   cybersecurity.                                                                                                                                                                                                                                                                                                                                                                                                                                                                                                                                                                                                                                                                                                                                                                                                                                                                                                                                                                                                                                            \\ \hline
15 & OES Incident Notification Form                                         & Operators of Essential Services:   Incident Notification Form                                                                                                                                           &      & Incident Notification Form to be   used by Operators of Essential Services                                                                                                                                                                                                                                                                                                                                                                                                                                                                                                                                                                                                                                                                                                                                                                                                                                                                                                                                                                                                                                                                                    \\ \hline
16 & The Department of the   Environment, Climate and Communications (DECC) & The Department of the   Environment, Climate and Communications                                                                                                                                         &      & The DECC creates   policies for Communications and Digital including Cybersecurity                                                                                                                                                                                                                                                                                                                                                                                                                                                                                                                                                                                                                                                                                                                                                                                                                                                                                                                                                                                                                                                                            \\ \hline
17 & The National Cyber Security Centre (NSCS)                              & The National Cyber Security   Centre                                                                                                                                                                    &      & The NCSC of Ireland  provides enhanced services to government   agencies and critical infrastructure providers to assist them in defending   against cyber security incidents, current threats and vulnerabilities   associated with network information security. The NCSC team works alongside   government agencies and private industry to facilitate secure systems and   information. NCSC participates in national, EU and international emergency   response exercises and provides expert advice to government departments and   industry on specific cyber threats and incidents.                                                                                                                                                                                                                                                                                                                                                                                                                                                                                                                                                                   \\ \hline
18 & CSIRT-IE                                                               & Computer Security Incident Response Team   (CSIRT-IE)                                                                                                                                                   &      & The CSIRT-IE is responsible of   the incident response for national cyber security incidents at the government   and national levels.                                                                                                                                                                                                                                                                                                                                                                                                                                                                                                                                                                                                                                                                                                                                                                                                                                                                                                                                                                                                                         \\ \hline
19 & Data Protection Commission (DPC)                                       & Data Protection Commission                                                                                                                                                                              &      & The DPC is the national   independent authority responsible for upholding the fundamental right of   individuals in the EU to have their personal data protected. The DPC is the   Irish supervisory authority for the General Data Protection Regulation   (GDPR), and also has functions and powers related to other important   regulatory frameworks including the Irish ePrivacy Regulations (2011) and the   EU Directive known as the Law Enforcement Directive.                                                                                                                                                                                                                                                                                                                                                                                                                                                                                                                                                                                                                                                                                       \\ \hline
20 & The European Union Agency for   Cybersecurity (ENISA)                  & The European Union Agency for   Cybersecurity                                                                                                                                                           &      & The European Union Agency for   Cybersecurity, ENISA, is the Union’s agency dedicated to achieving a high   common level of cybersecurity across Europe. Established in 2004 and   strengthened by the EU Cybersecurity Act, the European Union Agency for   Cybersecurity contributes to EU cyber policy, enhances the trustworthiness of   ICT products, services and processes with cybersecurity certification   schemes, cooperates with Member States and EU bodies, and helps Europe   prepare for the cyber challenges of tomorrow.                                                                                                                                                           
\\ \hline
21 & ISACA Cybersecurity Professional   Association                         & ISACA Cybersecurity Professional   Association                                                                                                                                                          &      & ISACA is an international   professional association focused on IT (information technology) governance.   It offers training and certification programs and it is the creator of the   COBIT framework. The ISACA Ireland Chapter's aim is to sponsor local educational   seminars and workshops, conduct regular chapter meetings, and help to further   promote and elevate the visibility of the IS audit, control and security   profession throughout the island of Ireland.                                                                                                                                                                                                             \\ \hline
22 & Be Safe Online - Ireland's   Official Online Safety Hub                & Be Safe Online                                                                                                                                                                                          &      & Be Safe Online is the   government’s campaign to highlight ways to help you stay safe online. The webpage provides access to a wide range of Online Safety resources, to   support online safety for all.                                                                                                            \\ \hline
23 & NIS Directive                                                          & DIRECTIVE (EU) 2016/1148 OF THE   EUROPEAN PARLIAMENT AND OF THE COUNCIL of 6 July 2016 concerning measures for   a high common level of security of network and information systems across the   Union &      &                                                                                                                                                                                                                                                                                                                                                                                                                                                                                                                                                                                                                                                                                                                                                                                                                                                                                                                                                                                                                                                                                                                                                               \\ \hline
24 & ENISA Guideline Incident Reporting                                     & ENISA Technical Guideline on   Incident Reporting under the EECC                                                                                                                                        &      &                                                                                                                                                                                                                                                                                                                                                                                                                                                                                                                                                                                                                                                                                                                                                                                                                                                                                                                                                                                                                                                                                                                                                               \\ \hline
25 & ENISA Guideline Security Measures                                      & ENISA Guideline on Security   Measures under the EECC                                                                                                                                                   &      &                                                                                                                                                                                                                                                                                                                                                                                                                                                                                                                                                                                                                                                                                                                                                                                                                                                                                                                                                                                                                                                                                                                                                               \\ \hline
26 & AI Cybersecurity Challenges                                            & AI Cybersecurity Challenges -   Threat Landscape for Artificial Intelligence                                                                                                                            & 2020 & This report presents the   Agency’s active mapping of the AI cybersecurity ecosystem and its Threat   Landscape, realised with the support of the Ad-Hoc Working Group on   Artificial Intelligence Cybersecurity.                                                                                                                                                                                                                                                                                                                                                                                                                                                                                                                                                                                                                                                                                                                                                                                                                                                                                                                                            \\ \hline
\end{longtable}
}

\newpage
\subsection{Qualitative Findings}

Table~\ref{tab:qualitative_findings} presents the qualitative findings of this study.

\footnotesize{
\begin{longtable}{ | m{1em} | m{11em}| m{3.5cm} |  m{6cm} |} 
\caption{Qualitative Findings}
\label{tab:qualitative_findings}
\endfirsthead
\hline 
{\cellcolor[RGB]{0,152,187}}\color{white}\textbf{No} & {\cellcolor[RGB]{0,152,187}}\color{white}\textbf{Qualitative Findings -   Questions} & {\cellcolor[RGB]{0,152,187}}\color{white}\textbf{Ireland Evidence} & {\cellcolor[RGB]{0,152,187}}\color{white}\textbf{Ireland Findings} 
 \\  \hline
\endhead
 \hline 
{\cellcolor[RGB]{0,152,187}}\color{white}\textbf{No} & {\cellcolor[RGB]{0,152,187}}\color{white}\textbf{Qualitative Findings -   Questions} & {\cellcolor[RGB]{0,152,187}}\color{white}\textbf{Ireland Evidence} & {\cellcolor[RGB]{0,152,187}}\color{white}\textbf{Ireland Findings} 
 
\\  \hline
\multirow{2}{*}{1} & \multirow{2}{16.5em}{What is the government cybersecurity policy unit?} & The Department of the   Environment, Climate and Communications (DECC) & The   DECC creates policies for Communications and Digital including Cybersecurity. \\ \cline{3-4} &    & The National Cyber Security Centre (NSCS) & The   NCSC of Ireland  provides enhanced   services to government agencies and critical infrastructure providers to   assist them in defending against cyber security incidents, current threats   and vulnerabilities associated with network information security. The NCSC   team works alongside government agencies and private industry to facilitate   secure systems and information. NCSC participates in national, EU and   international emergency response exercises and provides expert advice to   government departments and industry on specific cyber threats and incidents.
  
\\  \hline
2 & What   are the government committees, working groups, etc. for cybersecurity?                            & Data Protection   Commission (DPC) & The   DPC is the national independent authority responsible for upholding the   fundamental right of individuals in the EU to have their personal data   protected. The DPC is the Irish supervisory authority for the General Data   Protection Regulation (GDPR), and also has functions and powers related to   other important regulatory frameworks including the Irish ePrivacy   Regulations (2011) and the EU Directive known as the Law Enforcement   Directive.   

\\ \hline

3 & What is the government national-level cybersecurity strategy or other equivalent document? & National   CS Strategy &  The   National Cyber Security Strategy (December 2019) follows on from the   country's first strategy. It is a broader and more comprehensive document   than the last one, and is informed by the operational experience gained by   the National Cyber Security Centre (NCSC) from 2015 to 2019, and from ongoing   national and international engagements in the area.     The vision behind the 2019 Strategy is for Ireland to continue to safely   enjoy the benefits of the digital revolution and to play a full part in shaping the future of the Internet. This vision will be achieved   through: 

 \begin{itemize}
  \item the protection of the State, its people, and its critical national   infrastructure from cyber threats
  \item the development of the capacity of the State, of research institutions,   of businesses and of the people, to both better understand and manage the   nature of the challenges we face
  \item the engagement by the State, nationally and internationally, in a   strategic manner, supporting a free, open, peaceful and secure cyber space 
\end{itemize}

\\ \hline

\multirow{5}{*}{4}  & \multirow{5}{11.5em}{What is the government implementation plan to   the national-level cybersecurity strategy or other equivalent document?} & National CS   Strategy & Appendix   1 contains the list of actions for implementing the cybersecurity strategy.                                                                                                                

\\ \cline{3-4} & & Operators of Essential Service  & The   NIS Compliance Guidelines for Operators of Essential Service (OES)   establishes a set of Guidelines designed to assist OES in meeting their   network and information system security and incident reporting requirements   under Regulations 17 and 18 of the NIS Regulations. As part of the EU   Cybersecurity strategy the European Commission, the EU Network and   Information Security (NIS) Directive (see EU 2016/1148) is the first piece of   EU-wide cybersecurity legislation. The goal is to enhance cybersecurity across   the EU.                                                                                                           \\ \cline{3-4} & & Digital Service Providers & The   Information Note for Digital Service Providers assists Digital Service   Providers in understanding their obligations in relation to compliance with   the NIS Directive. As part of the EU Cybersecurity strategy the European   Commission, the EU Network and Information Security (NIS) Directive (see EU   2016/1148) is the first piece of EU-wide cybersecurity legislation. The goal   is to enhance cybersecurity across the EU.                                                          

\\ \cline{3-4} & & IS Management & The   Advice Note: Considering Information Security Management Supporting the   Public Service ICT Strategy contains the suggested strategies and activities   to enhance information security. t aims to assist Public Bodies in making   informed, risk-based decisions in relation to the improvement of Information   Security within their organisation.                                                                                                                                                         \\ \cline{3-4} & & CS Guidance for Business & The   12 Steps to Cyber Security Guidance on Cyber Security for Irish Business is   intended to be used by businesses as a suggested activity plan which may be   undertaken on a month-by-month basis over a suggested 12 month period to   improve cyber resilience.  

\\ \hline

5  & What   is the government entity unit that is specialised in national strategic cyber   threats?         & The National   Cyber Security Centre (NSCS) & The NCSC of Ireland  provides enhanced services to government   agencies and critical infrastructure providers to assist them in defending   against cyber security incidents, current threats and vulnerabilities   associated with network information security. The NCSC team works alongside   government agencies and private industry to facilitate secure systems and   information. NCSC participates in national, EU and international emergency   response exercises and provides expert advice to government departments and   industry on specific cyber threats and incidents.                                                                                                                           \\ \hline

\multirow{2}{*}{6}  & \multirow{2}{12em}{What are the cyber threat reports published by   the government?}   & \multirow{2}{11.5em}{The National Cyber Security   Centre (NSCS)} & The   National Cyber Security Centre (NCSC) is regularly publishing Alerts \&   Advisories on cyber security issues that may affect Ireland   (https://www.ncsc.gov.ie/news/) The latest published alerts are the   following: 

\begin{itemize}
  \item 11-03-2021 Critical Vulnerabilities in F5 products
  \item 10-03-2021 Critical Vulnerabilities in Microsoft Exchange Servers Update   2
  \item 25-02-2021 Unauthorised RCE in VMware vCenter \& ESXi
  \item 10-02-2021 Windows TCP/IP Remote Code Execution \& DoS Vulnerabilities
  \item 25-01-2021 SonicWall Vulnerability
\end{itemize}

\\ \cline{3-4} &    & The European   Union Agency for Cybersecurity (ENISA) & ENISA   is the Union’s agency dedicated to achieving a high common level of   cybersecurity across Europe. ENISA provides the reports EU on threats as   below:  - ENISA (Annual) Threat Landscape (ETL) Report provides an overview of   threats, together with current and emerging trends. For example, ENISA Threat   Landscape 2020 - List of top 15 threats.- ENISA Advisory Info Notes provides the information in relation to the   latest threats and vulnerabilities.                                                                                 
\\ \hline
\multirow{4}{*}{7}  & \multirow{4}{12em}{What are the government cyber safety and   security websites?}      & The Department   of the Environment, Climate and Communications (DECC) & It   creates policies for Communications and Digital including Cybersecurity   (https://www.gov.ie/en/policy-information/5e101b-network-and-information-security-cyber-security/)                                                              
\\ \cline{3-4} & & The National   Cyber Security Centre (NSCS) & It   is  the primary Cyber Security   authority in the State. The NCSC of Ireland    provides enhanced services to government agencies and critical   infrastructure providers to assist them in defending against cyber security   incidents, current threats and vulnerabilities associated with network   information security (https://www.ncsc.gov.ie/) 

\\ \cline{3-4} & & CSIRT-IE & It   is responsible of the incident response for national cyber security incidents   at the government and national levels (https://www.ncsc.gov.ie/CSIRT/)                                                                                                                       
\\ \cline{3-4} & & Data Protection Commission (DPC) & It   is is the national independent authority responsible for upholding the   fundamental right of individuals in the EU to have their personal data   protected. The DPC is the Irish supervisory authority for the General Data   Protection Regulation (GDPR). (https://www.dataprotection.ie)                                                                        

\\ \hline

\multirow{2}{*}{8}  & \multirow{2}{12em}{What is the training on cibersecurity offered by   government?}  & National CS   Strategy &  One   of the main objectives of the National Cybersecurity Strategy is to invest in   educational initiatives to prepare the workforce for advanced IT and   cybersecurity careers.  These are the   main cybersecurity programs at the national level:      Bachelor’s level cyber security programmes:

\begin{itemize}
  \item https://www.wit.ie/courses/bsc\_hons\_ in\_computer\_forensics\_and\_security
  \item https://www.itcarlow.ie/courses/type/ undergraduate-cao-courses/computing-networking-courses/bsc-science-cybercrime-itsecurity-cw227.htm
  \item https://www.tudublin.ie/study/ undergraduate/courses/computing-dig-forensics-and-cyber-sec-tu863/
\end{itemize}

Master’s level cyber security   programmes:      

\begin{itemize}
  \item http://www.itb.ie/StudyatITB/ bn528.html
  \item http://www.ucd.ie/cci/education /prospective\_students/msc\_difc.html
  \item https://www.griffith.ie/ faculties/computing/courses/master-science-network-and-info 
\end{itemize}

\\ \cline{3-4} &  & ISACA Cybersecurity Professional Association & ISACA   is an international professional association focused on IT (information   technology) governance. It offers training and certification programs and it   is the creator of the COBIT framework. The ISACA Ireland Chapter's aim is to   sponsor local educational seminars and workshops, conduct regular chapter   meetings, and help to further promote and elevate the visibility of the IS   audit, control and security profession throughout the island of   Ireland.(http://www.isaca.org/chapters5/ireland/)                                                                                                                                                            \\ \hline

9  & What   are the international cooperation activites of government dedicated to cyber   security  (e.g. FIRST)?  & CSIRT-IE & The CSIRT-IE is   member of FIRST. FIRST's objective is to bring together incident response and   security teams from every country across the world to ensure a safe internet   for all.(https://www.first.org/members/teams/csirt-ie)                                                                                                                                                                                                                                     \\ \hline

10  & What   are the government cybersecurity guidelines/notes for the digital service   providers?  & Digital Service   Providers                                            & The   Information Note for Digital Service Providers assists Digital Service   Providers (i.e., Online marketplaces, Online search engines, and Cloud   computing services) in understanding their obligations in relation to   compliance with the NIS Directive.                                                                                                                                                      
\\ \hline

11  & What   are the government cybersecurity guidelines/notes for operators of essential   services? & Operators of Essential Service  & The NIS Compliance Guidelines   for Operators of Essential Service (OES)    (including digital infrastructure) establishes a set of Guidelines   designed to assist OES in meeting their network and information system security   and incident reporting requirements under Regulations 17 and 18 of the NIS Regulations.                                                                                                                                                                                                                                                                                        
\\ \hline

\multirow{2}{*}{12} & \multirow{2}{11em}{What evidence of the effective implementation of   cyber/information security policies (e.g. audit result, documentation,   specific report) is provided by operators of essential services?} 

& \multirow{2}{12em}{Security Regulation and Law} & S.I. No. 360/2018 - European Union (Measures for   a High Common Level of Security of Network and Information Systems)   Regulations 2018      IMPLEMENTATION AND ENFORCEMENT      Security assessment  27. (1)(a) The competent authority referred to in Regulation 7(1) may, in   relation to those sectors in respect of which it is designated as the   competent authority, carry out an assessment, whether by means of a security   audit or otherwise, of the compliance by an operator of essential services   with its obligations under Regulations 17 and 18 and for that purpose may   appoint an independent person or auditor to carry out the assessment on its   behalf.      (b) The competent authority referred to in Regulation 8 may, in relation to   a relevant digital service provider, carry out an assessment, whether by   means of a security audit or otherwise, of the compliance by a relevant   digital service provider with its obligations under Regulations 21 and 22 and   for that purpose may appoint an independent person or auditor to carry out   the assessment on its behalf.     (2) A competent authority referred to in paragraph (1) may request an   operator of essential services or a relevant digital service provider, as the   case may be, to provide the competent authority with—      (a) the information necessary for that competent authority to assess the   security of the network and information systems of the operator or provider,   as the case may be, including documented security policies, and      (b) evidence of the effective implementation by the operator or provider,   as the case may be, of security policies including the implementation of any   recommendations made on foot of a security audit or other assessment.                                                                                                                                                                                                                             
\\ \hline

13  & What   are the government legal acts for personal data protection?  & Data Protection   Regulation  & From 25 May 2018   the key legislative frameworks are: - General Data Protection Regulation (GDPR) - Data Protection Act 2018 - the “Law Enforcement Directive” (Directive (EU) 2016/680) which has been   transposed into Irish law by way of the Data Protection Act 2018 - the Data Protection Acts 1988 and 2003 - the 2011 “ePrivacy Regulations” (S.I. No. 336 of 2011 – the European   Communities (Electronic Communications Networks and Services) (Privacy And   Electronic Communications) Regulations 2011) (https://www.dataprotection.ie/en/dpc-guidance/law/data-protection-legislation)                                                                                                                                                                                     \\ \hline

14 & What are the government units (CSIRT, CERT, CIRT, etc.)   specialised in national-level cyber incident detection and response? & {CSIRT-IE}   & The government has a unit (CSIRT, CERT, CIRT,   etc.) that is specialised in national-level cyber incident detection and   response. The CSIRT-IE is responsible of the incident response for national   cyber security incidents at the government and national levels.                                                                                                                                                                                \\ \hline

\multirow{3}{*}{15} & \multirow{3}{11em}{What are the government authorities that digital   service providers and operators of essential services must notify of   cybersecurity incidents?} & OES Incident   Notification Form & Operators   of Essential Services (OES) are required to report incidents which fall under   the scope of the NIS Directive. A reportable incident is any incident which   has a significant impact on the continuity of an essential service which an   Operator of Essential Services provides. OES have to notify the CSIRT of an   incident by using the "Operators of Essential Services: Incident   Notification Form".                                                                                    

\\ \cline{3-4}  &   & Incident Report Form & As stated in Article 16 (4) of the NIS   Directive, in order to determine whether the impact of an incident is   substantial, the following parameters in particular shall be taken into   account: 

\begin{itemize}
  \item The number of users affected by the incident, in particular users relying   on the service for the provision of their own services;
  \item The duration of the incident;
  \item The geographical spread with regard to the area affected by the   incident;
  \item The extent of the disruption of the functioning of the service;
  \item The extent of the impact on economic and societal activities. Digital service providers have to notify the CSIRT of an incident by by   using the "Digital Service Providers: Incident Notification Form".
\end{itemize}

\\ \hline
\multirow{4}{*}{16} & \multirow{4}{11em}{What is the government crisis management plan   for large-scale cyber incidents?} & Operators   of Essential Service & Appendix B:   Security Guidelines Category: Service Protection Policies, Processes and   Procedures (PR.SP) Subcategory PR.SP-9: Response plans (Incident Response and   Business Continuity) and recovery plans (Incident Recovery and Disaster   Recovery) are in place and managed Subcategory PR.SP-10: Response and   recovery plans are tested. 

\\ \cline{3-4}  &    & Digital Service Providers    & Section 3.1:   Article 16 (1) Business continuity management referred to in point (c) of   Article 16(1) of Directive (EU)2016/1148 means the capability of an   organisation to maintain or as appropriate restorethe delivery of services at acceptable predefined levels following a disruptive incident andshall   include: (a)the establishment and the use of contingency plans based on a business impactanalysis for ensuring the continuity of the services provided   by digital service providerswhich shall be assessed and tested on a regular   basis for example, through exercises; (b)disaster recovery capabilities which   shall be assessed and tested on a regular basisfor example, through   exercises.     

\\ \cline{3-4}  &  & IS Management & Appendix   C: Information Security Frameworks ISO 27002 (A.17) ENISA (D6).  
\\ \cline{3-4}  &  & CS Guidance for Business  & STEP   9. Adopt a risk-based approach to resilience.                                                                                       
\\ \hline

\multirow{2}{*}{17} & \multirow{2}{11em}{What are the cybersecurity standards referenced   by the government?} & Operators of   Essential Service     & Appendix B:   Security Guidelines

\begin{itemize}
  \item CIS CSC
  \item COBIT 5
  \item ISA 62443-2-1:2009
  \item ISA 62443-3-3:2013
  \item ISO/IEC 27001:2013
  \item NIST SP 800-53 Rev. 4

\end{itemize}

\\ \cline{3-4} &   & IS Management & Appendix   C: Information Security Frameworks

\begin{itemize}
  \item ISO/IEC 27001 and ISO/IEC 27K Series
  \item ENISA – Technical Guideline on Security Measures
  \item National Institute of Standards and Technology (NIST)
  \item NIST Risk Management Framework (RMF)
  \item NIST Cybersecurity Framework
  \item ISACA COBIT 5 for Information Security
\end{itemize}

\\ \hline

\multirow{5}{*}{18} & \multirow{5}{11em}{Is there any specific cybersecurity information   for emerging technologies (AI, ML, etc.), cloud services, and operational   technologies (OT) (e.g., asset management, OT devices)?}        & National CS   Strategy & Executive   Summary (some emerging technologies listed)                                                                                                                                                                                                                                                                         \\ \cline{3-4} & & AI Cybersecurity   Challenges  & This report   presents the Agency’s active mapping of the AI cybersecurity ecosystem and   its Threat Landscape, realised with the support of the Ad-Hoc Working Group   on Artificial Intelligence Cybersecurity.                                                                                                                                                                                   \\ \cline{3-4}  &    & IS Management      & Section: Other   Useful Sources.      
\\ \cline{3-4}  &    & CS Guidance for Business    & Section:   Background (IoT mentioned)    
\\ \cline{3-4}  &    & AmCham GCS Report    & The main trends in the 2018   cyberthreat landscape, as identified by the European Union Agency for Network   and Information Security (ENISA) 19 showcase a range of hot-button issues.                                                                                                                                                      
\\ \hline

\end{longtable}
}

\pagebreak
\subsection{Capability Maturity Results}

Table~\ref{tab:maturity_results} presents the results of the maturity assessment of Ireland in this study.

\footnotesize{
\begin{longtable}{ | m{1em} | m{11em}| m{3cm} |  m{5cm} | m{1.2cm} |} 
\caption{Capability Maturity Results}
\label{tab:maturity_results}

\endfirsthead

\hline 
{\cellcolor[RGB]{0,152,187}}\color{white}\textbf{No} & {\cellcolor[RGB]{0,152,187}}\color{white}\textbf{Questions} & {\cellcolor[RGB]{0,152,187}}\color{white}\textbf{Evidence} & {\cellcolor[RGB]{0,152,187}}\color{white}\textbf{Findings} & {\cellcolor[RGB]{0,152,187}}\color{white}\textbf{Maturity Level} 
\\  \hline

\endhead
\hline 
{\cellcolor[RGB]{0,152,187}}\color{white}\textbf{No} & {\cellcolor[RGB]{0,152,187}}\color{white}\textbf{Questions} & {\cellcolor[RGB]{0,152,187}}\color{white}\textbf{Evidence} & {\cellcolor[RGB]{0,152,187}}\color{white}\textbf{Findings} & {\cellcolor[RGB]{0,152,187}}\color{white}\textbf{Maturity Level} 
\\  \hline

\multirow{4}{*}{1}   & \multirow{4}{12em}{To what extent are   policies, laws, and guidelines for information security related to VTI\&CM   established?} & National CS   Strategy  & The vision   behind the 2019 Strategy is for Ireland to continue to safely enjoy the   benefits of the digital revolution and to play a full part in shaping the   future of the Internet. This vision will be achieved through:

\begin{itemize}
  \item the protection of the State, its people, and its critical national   infrastructure from cyber threats
  \item the development of the capacity of the State, of research institutions,   of businesses and of the people, to both better understand and manage the   nature of the challenges we face
  \item the engagement by the State, nationally and internationally, in a   strategic manner, supporting a free, open, peaceful and secure cyber space

\end{itemize}

& \multirow{4}{*}{3}   

\\ \cline{3-4} &     & Operators of Essential Service    & The   NIS Compliance Guidelines for Operators of Essential Service (OES)   establishes a set of Guidelines designed to assist OES in meeting their   network and information system security and incident reporting requirements   under Regulations 17 and 18 of the NIS Regulations. &

\\ \cline{3-4} &  & Digital Service Providers      & The   Information Note for Digital Service Providers assists Digital Service   Providers in understanding their obligations in relation to compliance with   the NIS Directive. &                                 

\\ \cline{3-4} &  & CS Guidance for Business  & Ireland   has developed guidelines for businesses to develop/adopt policies and   standards (security policy, data protection policy, malware protection   policy, remote access policy, etc.), as well as a programme to  monitor/measure compliance.  &   

\\ \hline \pagebreak \hline

\multirow{5}{*}{2}   & \multirow{5}{12em}{To   what extent are there guidelines and regulations for development and   execution of information security strategies established?} & National   CS Strategy                                & The   Appendix 1 of the National Cybersecurity Strategy contains the list of actions for   implementing the sybersecurity strategy. 

& \multirow{5}{*}{3} \\ \cline{3-4} &    & Operators of Essential Service  & The   NIS Compliance Guidelines for Operators of Essential Service (OES)   establishes a set of Guidelines designed to assist OES in meeting their   network and information system security and incident reporting requirements   under Regulations 17 and 18 of the NIS Regulations. &   

\\ \cline{3-4} &   & Digital Service Providers  & The   Information Note for Digital Service Providers assists Digital Service   Providers in understanding their obligations in relation to compliance with   the NIS Directive    & 

\\ \cline{3-4}  &  & IS Management  & The   Advice Note: Considering Information Security Management Supporting the Public Service ICT Strategy contains the suggested   strategies and activities to enhance information security &

\\ \cline{3-4} &  & CS Guidance for Business & The 12 Steps to Cyber Security Guidance on Cyber Security for Irish Business is   intended to be used by businesses as  a   suggested activity plan which may be undertaken on a month-by-month basis over a suggested 12 month period to improve cyber   resilience &   

\\ \hline

\multirow{2}{*}{3}   & \multirow{2}{12em}{To   what extent are governance structures for    information security related to VTI\&CM established?}                                & National   CS Strategy                                & The National Cyber Security Centre (NCSC)   will remain the primary Cyber Security authority in the State. The NCSC of   Ireland  provides enhanced services to   government agencies and critical infrastructure providers to assist them in   defending against cyber security incidents, current threats and   vulnerabilities associated with network information security. The NCSC team   works alongside government agencies and private industry to facilitate secure   systems and information. NCSC participates in national, EU and international   emergency response exercises and provides expert advice to government   departments and industry on specific cyber threats and incidents                                                                                                                                                                                                                                                                                                                                                                                                                                                                    & \multirow{2}{*}{3}              \\ \cline{3-4}
                     &                                                                                                                                                        & Requirements for CSIRTs                               & The   government has a unit (CSIRT, CERT, CIRT, etc.) that is specialised in   national-level cyber incident detection and response. The CSIRT-IE is   responsible of the incident response for national cyber security incidents at   the government and national levels.                                                                                                                                                                                                                                                                                                                                                                                                                                                                                                                                                                                                                                                                                                                                                                                                                                                                                                 &                                 \\ \hline
\multirow{4}{*}{4}   & \multirow{4}{12em}{To what extent are responsibilities, penalties,   incentive models, and audit options established?}                                    & National   CS Strategy                                & The   NCSC has developed significantly in terms of capacity and resources, and its   roles have been formally established in law, including responsibilities   around Critical National Infrastructure protection and dealing with EU   requirements around the security of some. This document also focuses on the   ongoing compliance and audit programmes to mitigate risks to key services.                                                                                                                                                                                                                                                                                                                                                                                                                                                                                                                                                                                                                                                                                                                                                                           & \multirow{4}{*}{3}              \\ \cline{3-4}
                     &                                                                                                                                                        & Requirements for CSIRTs                               & The   status of CSIRT-IE and its roles and functions are defined and documented.   The requirement CSIRT-IE RFC2350 provides basic information about a CSIRT   including contact details and its roles and responsibilities.                                                                                                                                                                                                                                                                                                                                                                                                                                                                                                                                                                                                                                                                                                                                                                                                                                                                                                                                               &                                 \\ \cline{3-4}
                     &                                                                                                                                                        & Digital Service Providers                             & Defines   responsibilities for digital service providers.                                                                                                                                                                                                                                                                                                                                                                                                                                                                                                                                                                                                                                                                                                                                                                                                                                                                                                                                                                                                                                                                                                                  &                                 \\ \cline{3-4}
                     &                                                                                                                                                        & Operators of Essential Service                        & Presents   guidelines to inform about cyber security roles and responsibilities for   operators of essential services. (Detail standards)      &                                

 \\ \hline
 
\multirow{4}{*}{5} & \multirow{4}{12em}{What approaches are in place for threat   profiling and managing information security risks and vulnerabilities?}   & Operators of   Essential Service & Secion A. Identify:     

\begin{itemize}
  \item v. Risk Management Strategy
  \item vi. Supply Chain Risk Management
\end{itemize}

Appendix B: Security Guidelines. Categories:
\begin{itemize}
  \item Risk Management Strategy (ID.RM)
  \item Supply Chain Risk Management (ID.SC)
\end{itemize}

Subcategories:
\begin{itemize}
  \item PR.SP-12: A vulnerability management plan is developed and implemented to   remediate vulnerabilities in a timely manner, commensurate with the risk 
\end{itemize}

& \multirow{4}{*}{3} \\ \cline{3-4} &  & IS Management   &  

\begin{itemize}
  \item ISMS   Focus Area 2 - Adopt a Risk Management Approach
  \item Appendix C: Information Security Frameworks
  \item NIST Risk Management Framework (RMF)
  \item ENISA (D1) - Governance and risk management
  \item ISO 27002 (A.12.6) - Technical vulnerability management
\end{itemize}

&  \\ \cline{3-4}  &  & CS Guidance for Business   & 
\begin{itemize}
  \item STEP   4. Define your risk appetite
  \item STEP 6. Implement basic protections (vulnerability management)
  \item STEP 12. Create a cyber risk management lifecycle
\end{itemize}

&  \\ \cline{3-4} &  & The National Cyber Security Centre (NSCS)     & The   National Cyber Security Centre (NCSC) is regularly publishing Alerts \&   Advisories on cyber security issues that may affect Ireland   (https://www.ncsc.gov.ie/news/) &     

\\ \hline \multirow{3}{*}{6}   & \multirow{3}{12em}{To what extent is information security   management training developed and disseminated?}                                              & National   CS Strategy                                & Skillnet   Ireland launched their Cyber Security Skills Initiative to deliver a broad   programme of initiatives in the field. Government has launched Future Jobs   Ireland, a multiannual framework for skills and enterprise development,   including the technology sector. Ireland offers a significant number of   courses in cyber security, with at least 8 Masters level courses now on   offer   

& \multirow{3}{*}{4}   \\ \cline{3-4} &     &   & NATO  &   

\\ \cline{3-4}  &   & Operators of Essential Service    & All   users are informed and trained on cyber security policies and relevant   procedures, with periodic updates. (Detail standards)   

&                                 \\ \hline
\multirow{3}{*}{7}   & \multirow{3}{12em}{To what extent is information security   management training developed and disseminated?}                                              & Incident   notification for DSPs                      & Incident   notification for DSPs in the context of the NIS Directive: As part of the EU   Cybersecurity strategy the European Commission proposed the EU Network and   Information Security directive. The NIS Directive (see EU 2016/1148) is the   first piece of EU-wide cybersecurity legislation. IS imposes security   measures and incident reporting obligations for Operators of Essential   Services (OES) and Digital Service Providers (DSP).                                                                                                                                                                                                                                                                                                                                                                                                                                                                                                                                                                                                                                                                                                                  & \multirow{3}{*}{3}              \\ \cline{3-4}
                     &                                                                                                                                                        & National CS Strategy                                  & Details   implementation plan of actions relating to these measures - Information Note   for Digital Service Providers                                                                                                                                                                                                                                                                                                                                                                                                                                                                                                                                                                                                                                                                                                                                                                                                                                                                                                                                                                                                                                                     &                                 \\ \cline{3-4}
                     &                                                                                                                                                        & Operators of Essential Service                        &                                                                                                                                                                                                                                                                                                                                                                                                                                                                                                                                                                                                                                                                                                                                                                                                                                                                                                                                                                                                                                                                                                                                                                            &                                 \\ \hline
\multirow{3}{*}{8}   & \multirow{3}{12em}{To what extent are implementation frameworks   developed and standards adopted?}                                                       & National   CS Strategy                                & Complete   first phase of Operators of Essential Services (OES) Self-Assessment against   Security Control Framework                                                                                                                                                                                                                                                                                                                                                                                                                                                                                                                                                                                                                                                                                                                                                                                                                                                                                                                                                                                                                                                       & \multirow{3}{*}{3}              \\ \cline{3-4}
                     &                                                                                                                                                        & Operators of Essential Service                        & Appendix A: Framework   Infographics Appendix B: Security Guidelines                                                                                                                                                                                                                                                                                                                                                                                                                                                                                                                                                                                                                                                                                                                                                                                                                                                                                                                                                                                                                                                                                                       &                                 \\ \cline{3-4}
                     &                                                                                                                                                        & IS Management                                         & Appendix C:   Information Security Frameworks                                                                                                                                                                                                                                                                                                                                                                                                                                                                                                                                                                                                                                                                                                                                                                                                                                                                                                                                                                                                                                                                                                                              &                                 \\ \hline
\multirow{3}{*}{9}   & \multirow{3}{12em}{To   what extent are risks and challenges for cybersecurity implementation plans   identified and addressed?}                          & National   CS Strategy                                & National   Cybersecurity Strategy 2019 Cybersecurity challenges described in general                                                                                                                                                                                                                                                                                                                                                                                                                                                                                                                                                                                                                                                                                                                                                                                                                                                                                                                                                                                                                                                                                       & \multirow{3}{*}{3}              \\ \cline{3-4}
                     &                                                                                                                                                        & IS Management                                         & Appendix   D: Sources of Information in Relation to Threats and Vulnerabilities                                                                                                                                                                                                                                                                                                                                                                                                                                                                                                                                                                                                                                                                                                                                                                                                                                                                                                                                                                                                                                                                                            &                                 \\ \cline{3-4}
                     &                                                                                                                                                        & AmCham GCS Report                                     & The   American Chamber recommends addressing cyber security challenges through the   Disruptive Technologies Innovation Fund which would meet societal, in   addition to commercial, priorities and enhance Ireland’s global research   reputation Resourcing of teams and retention of talent is a challenge for An   Garda Síochána and the Defence Forces, similar to private sector experiences   in this regard.                                                                                                                                                                                                                                                                                                                                                                                                                                                                                                                                                                                                                                                                                                                                                      &                                 \\ \hline
\multirow{5}{*}{10}  & \multirow{5}{11.5em}{To what extent are the emerging technologies and   future trends for VTI\&CM identified?}                                              & National CS   Strategy                                & Executive   Summary (some emerging technologies listed) 

& \multirow{5}{*}{3} \\ \cline{3-4}  &   & AI Cybersecurity   Challenges   & AI   may expose individuals and organizations to new, and sometimes unpredictable,   risks, and it may open new avenues in attack methods and techniques, as well   as creating new data protection challenges, especially in safety-critical   deployments such as in autonomous vehicles, smart manufacturing, eHealth,   etc.     In this document, Chapter 4 introduces the threat taxonomy of AI systems,   where relevant threats are presented and mapped to corresponding   assets:

\begin{itemize}
  \item Data
  \item Model
  \item Actors
  \item Processes
  \item Environment/Tools
  \item Artefacts
\end{itemize}

&                                 \\ \cline{3-4}
                     &                                                                                                                                                        & IS Management                                         & Other   Useful Sources                                                                                                                                                                                                                                                                                                                                                                                                                                                                                                                                                                                                                                                                                                                                                                                                                                                                                                                                                                                                                                                                                                                                                     &                                 \\ \cline{3-4}
                     &                                                                                                                                                        & CS Guidance for Business                              & Background   (IoT mentioned)                                                                                                                                                                                                                                                                                                                                                                                                                                                                                                                                                                                                                                                                                                                                                                                                                                                                                                                                                                                                                                                                                                                                               &                                 \\ \cline{3-4}
                     &                                                                                                                                                        & AmCham GCS Report                                     & The   main trends in the 2018 cyberthreat landscape, as identified by the European   Union Agency for Network and Information Security (ENISA) 19 showcase a range   of hot-button issues     &
                     
\\ \hline

\multirow{3}{*}{11}  & \multirow{3}{12em}{How does the government define security   architectures guidelines  and tools for   VTI\&CM?}                                          & IS Management                                         & Suggests   strategies and activities to enhance the Information Security Posture for   improving governance around ICT in the Public Service. The security posture   is defined as the security status of an enterprise’s networks, information,   and systems based on information architecture resources (e.g. people,   hardware, software, policies) and capabilities in place to manage the defence   of the enterprise and to react as the situation changes.                                                                                                                                                                                                                                                                                                                                                                                                                                                                                                                                                                                                                                                                                                        & \multirow{3}{*}{2}              \\ \cline{3-4}
                     &                                                                                                                                                        & CS   Guidance for Business                            & organisations   should consider additional protections to further reduce the risk of cyberattacks.   such as echnologies which support Access Management (IAM) and Vulnerability   Management (VM), and in particular the centralisation and automation of IAM   and VM processes will deliver efficiencies and cost savings. Consideration   should be given to establishing a formal role/function which considers   security architecture in all aspects of your organisation’s IT programme. It   is particularly important for organisations who develop software/systems that   security is embedded at all stages throughout the Systems Development   Lifecycle. Recommendation baed on NIST Special Publication 800-39 Managing Information   Security Risk.                                                                                                                                                                                                                                                                                                                                       &                                 \\ \cline{3-4}
                     &                                                                                                                                                        & Digital   Service Providers                           & The systematic   management of network and information systems which means amapping of   information systems and the establishment of a set of appropriate policieson   managing information security, including risk analysis, human resources,   security ofoperations, security architecture, secure data and system   lifecycle management andwhere applicable, encryption and its management.                                                                                                                                                                                                                                                                                                                                                                                                                                                                                                                                                                                                                                                                                                                                                                         &                                 \\ \hline
\multirow{5}{*}{12}  & \multirow{5}{12em}{To what extent are documents to define,   implement, and monitor measures to protect IT devices established?}                          & CS Asset Management Report                            & The report   highlights that devices that are exposed to a high number of known   vulnerabilities for a lengthy period of time are more vulnerable to malicious   actors who may gain unauthorised access to IT assets and compromise the   confidentiality, integrity and availability of stored business critical data.                                                                                                                                                                                                                                                                                                                                                                                                                                                                                                                                                                                                                                                                                                                                                                                                                                                  & \multirow{5}{*}{3}              \\ \cline{3-4}
                     &                                                                                                                                                        & IS Management                                         & Provides   the  guidelines  and recommendations for Information   Security Management  based on standards   such as the ISO 27002 Security Control Categories and Control Objectives   which specifies the control description for mobile devices and teleworking.                                                                                                                                                                                                                                                                                                                                                                                                                                                                                                                                                                                                                                                                                                                                                                                                                                                                                                         &                                 \\ \cline{3-4}
                     &                                                                                                                                                        & CS Guidance for Business                              & Recommends   to establish comprehensive anti-malware protection, both at the host (on   every computer,  server, mobile device,   etc.) and network-level (at email/web gateways, etc.) as a basic protection   for Irish busines. Additionally, it recommends establishing remote   access/remote working controls, including secure access to systems and   controls on the devices which may be used to access internal systems and   using full disk encryption on all mobile devices, such as laptops and   removable media (USB keys).                                                                                                                                                                                                                                                                                                                                                                                                                                                                                                                                                                                                                               &                                 \\ \cline{3-4}
                     &                                                                                                                                                        & National CS Strategy                                  & The   progressive dependence of Critical National Infrastructure and services on   network-connected devices has led to the State taking a series of measures to   ensure the resilience of certain categories of critical national   infrastructure. This strategy provides a comprehensive section (6) on   Critical National Infrastructure Protection and related measures.                                                                                                                                                                                                                                                                                                                                                                                                                                                                                                                                                                                                                                                                                                                                                                                            &                                 
\\ \cline{3-4} & & Operators of Essential Service   & This   guideline presents the measures for Asset Management, including data,   personnel, devices, systems, and facilities. Please see Appendix B: Security   Guidelines for recommended practices and standards to develop and implement   the appropriate and proportionate security measures:
                     
\begin{itemize}
  \item Asset Management (ID.AM)
  \item Identity Management, Authentication and Access Control (PR.AC)
\end{itemize}  &   

\\ \hline
\multirow{7}{*}{13}  & \multirow{7}{12em}{What information about threats is available at   the country level?}                                                                   & IS Management                                         & 

Appendix D   offers some potential information sources to stay abreast of the latest   threats, vulnerabilities, and potential security control options. This   includes:
\begin{itemize}
  \item ENISA (Annual) Threat Landscape (ETL) Report provides an overview of threats,   together with current and emerging trends
  \item ENISA Advisory Info Notes provides the information in relation to the   latest threats and vulnerabilities
\end{itemize} 

& \multirow{7}{*}{4}              \\ \cline{3-4}
                     &                                                                                                                                                        & CS Guidance for Business                              & This   guidance provides clear steps which organisations should include to   understand who might want to attack them, why, and how they might go about   carrying out such an attack in order to allow organisations to focus the   efforts on how to respond to the most likely threats. For example, it   recommends to establish a Cyber Threat Intelligence (CTI) capability which   enables organisations to identify (through intelligence sources/feeds) and   understand the top 5-10 threat actors and likely attack scenarios (these are   your key cyber risks) and record them in a risk register. Please, see section   3. Understand the Threats.                                                                                                                                                                                                                                                                                                                                                                                                                                                                                                           &                                 \\ \cline{3-4}
                     &                                                                                                                                                        & Incident Report Form                                  & This   incident report form includes a section of lessons learned based on new   threats identified.                                                                                                                                                                                                                                                                                                                                                                                                                                                                                                                                                                                                                                                                                                                                                                                                                                                                                                                                                                                                                                                                       &                                 
 \\ \cline{3-4}  &   & National CS Strategy                                  & 
 
 \begin{itemize}
  \item Expanding the current Threat Sharing Group (TSG) which was established in   2017. The TSG acts both as a forum for critical national infrastructure   operators, and a means for State Actors (including Gardaí and Defence Forces)   to share information with these operators and to engage with cybersecurity   professionals
  \item The NCSC has developed a threat intelligence database that is being used   to assist Agencies and Departments in protecting their networks
  \item The CSIRT has a proactive position that includes the deployment and use   of MISPs (Malware Information Sharing Platform) to share threat intelligence   directly with Critical National Infrastructure Providers, and the evolution   and use of a series of tools to identify, parse and analyse open-source   intelligence (OSINT). The CSIRT has also developed, tested, and deployed the   ‘Sensor’ platform, now operational on the infrastructure of a number of   Government Departments, to detect and warn of certain types of threat. 
\end{itemize}  &                                

\\ \cline{3-4}  &  & The National Cyber Security Centre (NSCS)             &

The   National Cyber Security Centre (NCSC) is regularly publishing Alerts \&   Advisories on cyber security issues that may affect Ireland   (https://www.ncsc.gov.ie/news/) The latest published alerts are the   following: 

\begin{itemize}
  \item 11-03-2021 Critical Vulnerabilities in F5 products
  \item 10-03-2021 Critical Vulnerabilities in Microsoft Exchange Servers Update   2 
  \item 25-02-2021 Unauthorised RCE in VMware vCenter \& ESXi
  \item 10-02-2021 Windows TCP/IP Remote Code Execution \& DoS Vulnerabilities 
  \item 25-01-2021 SonicWall Vulnerability
\end{itemize}  &

\\ \cline{3-4}  &    & The European Union Agency for Cybersecurity   (ENISA) & ENISA   is the Union’s agency dedicated to achieving a high common level of   cybersecurity across Europe. ENISA provides the reports EU on threats as   below:

\begin{itemize}
  \item ENISA (Annual) Threat Landscape (ETL) Report provides an overview of   threats, together with current and emerging trends. For example, ENISA Threat   Landscape 2020
  \item List of top 15 threats 
  \item ENISA Advisory Info Notes provides the information in relation to the   latest threats and vulnerabilities
\end{itemize} &                                 

\\ \cline{3-4}
                     &                                                                                                                                                        & Operators of Essential Service                        & Appendix   B outlines the best practices and informative references recommended by the   government for Operators of Essential Service in order to manage cyber threat   information.       &                                 \\ \hline
\multirow{4}{*}{14}  & \multirow{4}{12em}{What information about threats actors is   available at the country level?}    & CS Guidance for Business   & This guidance   provides clear steps which organisations should follow to defend their   business against threat actors. Please, refer to the following   sections:

\begin{itemize}
  \item 3. Understand the threats: Threat actors (cyber criminals, malicious   insiders, etc.) vary in capability and sophistication, whilst also constantly   changing depending on the value of the prize they seek to exploit.
  \item  5. Focus on education and awareness Establish an education and awareness   programme, ensuring all of your employees, contractors and third parties can   identify a cyberattack and are aware of the role they play in defending your   business against threat actors
  \item 7. Be able to detect an attack: Threat actors are many and sophisticated,   and dedicated attackers have a high chance of breaching organisations'   defences given enough time and persistence
\end{itemize} 

& \multirow{4}{*}{4}              \\ \cline{3-4}
                     &                                                                                                                                                        & National CS Strategy                                  & This   document identifies key threat actors such as those state-sponsored entities,   usually military or security organisations, seeking to use network and   information systems to conduct operations ranging from the exfiltration of   data to the destruction of physical infrastructure. These threat actors,   usually referred to as ‘advanced persistent threats’ (or APTs) have been   shown to be involved in attacks across a wide range of sectors, but with a   particular focus on Government IT systems, telecommunications networks,   financial services, and technology companies.  &                              
\\ \cline{3-4}  &   & The National Cyber Security Centre (NSCS)    & The   National Cyber Security Centre (NCSC) is regularly publishing Alerts \&   Advisories on cyber security issues that may affect Ireland   (https://www.ncsc.gov.ie/news/) The latest published alerts are the   following: 

\begin{itemize}
  \item 11-03-2021 Critical Vulnerabilities in F5 products
  \item 10-03-2021 Critical Vulnerabilities in Microsoft Exchange Servers Update   2
  \item 25-02-2021 Unauthorised RCE in VMware vCenter \& ESXi
  \item 10-02-2021 Windows TCP/IP Remote Code Execution \& DoS Vulnerabilities
  \item 25-01-2021 SonicWall Vulnerability 
\end{itemize} &                                

\\ \cline{3-4} &  & The European Union Agency for Cybersecurity (ENISA)   & ENISA   is the Union’s agency dedicated to achieving a high common level of   cybersecurity across Europe. ENISA provides the reports EU on threats as   below:

\begin{itemize}
  \item ENISA (Annual) Threat Landscape (ETL) Report provides an overview of   threats, together with current and emerging trends. For example, ENISA Threat   Landscape 2020 
  \item List of top 15 threats
  \item ENISA Advisory Info Notes provides the information in relation to the   latest threats and vulnerabilities
\end{itemize} &   

\\ \hline 

\multirow{5}{*}{15}  & \multirow{5}{12em}{To what extent are data security classification   guidelines established?}                                                             & CS Asset   Management Report                          & This document   outlines key findings in Irish Asset Management Firms. For example,   deficiencies in IT asset inventories were identified, where the inventories   did not capture the complete IT estate and / or classify assets (e.g., data   assets)  by their business criticality.                                                                                                                                                                                                                                                                                                                                                                                                                                                                                                                                                                                                                                                                                                                                                                                                                                                                                  & \multirow{5}{*}{3}              \\ \cline{3-4}
                     &                                                                                                                                                        & IS Management                                         & This   document recommends the ISO 27002 Security Control Categories and Control   Objectives which outlines the Information classification as part of the Asset   management.   &                                
\\ \cline{3-4} &   & CS Guidance for Business  & This   guide recommends that organisations should classify assets (e.g., data).   Please see section 2. Identify what matters most: Establish a comprehensive technology asset management  programme which includes all of the assets   above and ranks/classifies them by criticality to your business in a   centralised asset inventory.                                                                                                                                                                                                                                                                                                                                                                                                                                                                                                                                                                                                                                                                                                                                                                               &                                 \\ \cline{3-4}
                     &                                                                                                                                                        & Digital Service Providers                             & This   document recommends mapping secure data to appropriate policies on managing   information security. Please see 3.1 Article 16 (1).                                                                                                                                                                                                                                                                                                                                                                                                                                                                                                                                                                                                                                                                                                                                                                                                                                                                                                                                                                                                                                  &                                 \\ \cline{3-4}
                     &                                                                                                                                                        & Operators of Essential Service                        & This guideline   presents the measures for Asset Management, including data, personnel,   devices, systems, and facilities. Please see Appendix B: Security Guidelines   which outlines the best practices for data classification as part of the   Asset Management (ID.AM)  &                                
\\ \hline

\multirow{2}{*}{16}  & \multirow{2}{11.7em}{To what extent is data life cycle management   addressed by the government?}  & CS Guidance for   Business  & This guide   recommends the management of data lifecysle. Please see section 6. - Implement basic   protections. Establish a data security programme, which focuses on how data is stored   and secured in organisations. This involves mapping of data locations and   flows at step two above, and then focusing on how it is protected at every   step in its lifecycle.                                                                                                                                                                                                                                                                                                                                                                                                                                                                                                                                                                                                                                                                                                                                                  & \multirow{2}{*}{2}              \\ \cline{3-4}
                     &                                                                                                                                                        & Digital Service Providers                             & This   document recommends mapping information systems and the establishment of   appropriate policies on managing information security, including secure data   and system lifecycle management. Please see 3.1 Article 16 (1).                                                                                                                                                                                                                                                                                                                                                                                                                                                                                                                                                                                                                                                                                                                                                                                                                                                                                                                                           &                                 \\ \hline
17                   & To what extent   is data security administration addressed by the government?   & Data Protection   Regulation  & 

The Data Protection Commission   (DPC) is the national independent authority responsible for upholding the   fundamental right of individuals in the EU to have their personal data   protected. The DPC is the Irish supervisory authority for the General Data   Protection Regulation (GDPR). (https://www.dataprotection.ie/). 

\begin{itemize}
  \item There is comprehensive data protection legislation (up to date as of   31.01.2019).   (https://www.dataprotection.ie /en/dpc-guidance/law/data-protection-legislation)
  \item There are guidelines and checklists for organizations (https://www.dataprotection.ie/ en/organisations)
  \item There is information on own rights for individuals  (https://www.dataprotection.ie /en/individuals, https://www.citizensinformation. ie/en/government\_in\_ireland/ national\_government/standards \_and\_accountability/ data\_protection.html\#)
  \item There are up-to-date reports   (https://www.dataprotection.ie /en/dpc-guidance/publications). It is possible to report a breach or raise a complaint   (https://www.dataprotection.ie/) 
\end{itemize}

& 3                

\\ \hline
\multirow{5}{*}{18}  & \multirow{5}{11.5em}{To   what extent are identity and data access management addressed by the   government?}                                               & National CS   Strategy                                & Section   7.3: Measures 8. The NCSC will develop a baseline security standard to be   applied by all Government Departments and key agencies. These standards   typically include measures and controls in relation to staff training,   identity and access management.                                                                                                                                                                                                                                                                                                                                                                                                                                                                                                                                                                                                                                                                                                                                                                                                                                                                                                   & \multirow{5}{*}{3}              \\ \cline{3-4}
                     &                                                                                                                                                        & Operators of Essential Service                        & Section   4 B: Protect i. Identity Management, Authentication and Access Control                                                                                                                                                                                                                                                                                                                                                                                                                                                                                                                                                                                                                                                                                                                                                                                                                                                                                                                                                                                                                                                                                           &                                 \\ \cline{3-4}
                     &                                                                                                                                                        & Digital Service Providers                             & Section   3.1: Article 16 (1) (d)the access controls to network and information systems   which means the availability of a set of measures to ensure that the physical   and logical access to network and information systems, including   administrative security of network and information systems, is authorized and   restricted based on business and security requirements.                                                                                                                                                                                                                                                                                                                                                                                                                                                                                                                                                                                                                                                                                                                                                                                       &                                 \\ \cline{3-4}
                     &                                                                                                                                                        & IS Management                                         & Appendix   C: Information Security Frameworks ISO 27002 (A.9) - Access Control                                                                                                                                                                                                                                                                                                                                                                                                                                                                                                                                                                                                                                                                                                                                                                                                                                                                                                                                                                                                                                                                                             &                                 \\ \cline{3-4}
                     &                                                                                                                                                        & IS Management                                         & STEP   6. Implement basic protections                                                                                                                                                                                                                                                                                                                                                                                                                                                                                                                                                                                                                                                                                                                                                                                                                                                                                                                                                                                                                                                                                                                                      &                                 \\ \hline
\multirow{4}{*}{19}  & \multirow{4}{11.6em}{What   information security guidance is provided to support business continuity   planning and management?}                            & Operators   of Essential Service                      & Appendix   B: Security Guidelines Category: Service Protection Policies, Processes and   Procedures (PR.SP) Subcategory PR.SP-9: Response plans (Incident Response and   Business Continuity) and recovery plans (Incident Recovery and Disaster   Recovery) are in place and managed Subcategory PR.SP-10: Response and   recovery plans are tested                                                                                                                                                                                                                                                                                                                                                                                                                                                                                                                                                                                                                                                                                                                                                                                                                       & \multirow{4}{*}{3}              \\ \cline{3-4}
                     &                                                                                                                                                        & Digital Service Providers                             & Section   3.1: Article 16 (1) Business continuity management referred to in point (c)   of Article 16(1) of Directive (EU)2016/1148 means the capability of an   organisation to maintain or as appropriate restore the delivery of services   at acceptable predefined levels following a disruptive incident and shall   include: (a)the establishment and the use of contingency plans based on a   business impact analysis for ensuring the continuity of the services provided   by digital service providers which shall be assessed and tested on a regular   basis for example, through exercises; (b)disaster recovery capabilities which   shall be assessed and tested on a regular basis for example, through   exercises.                                                                                                                                                                                                                                                                                                                                                                                                                                    &                                 \\ \cline{3-4}
                     &                                                                                                                                                        & IS Management                                         & Appendix   C: Information Security Frameworks ISO 27002 (A.17) ENISA (D6)                                                                                                                                                                                                                                                                                                                                                                                                                                                                                                                                                                                                                                                                                                                                                                                                                                                                                                                                                                                                                                                                                                  &                                 \\ \cline{3-4}
                     &                                                                                                                                                        & CS Guidance for Business                              & STEP   9. Adopt a risk-based approach to resilience  &                                 
                     
\\ \hline

\multirow{5}{*}{20}  & \multirow{5}{12em}{To   what extent is security incidents and near incidents management addressed by   the government?}  & IS Management  & Appendix C:

\begin{itemize}
  \item Information Security Frameworks
  \item ISO 27002 (A.16) - Information security incident management
  \item ENISA (D5) - Incident management
\end{itemize}

& \multirow{5}{*}{3}              \\ \cline{3-4}
                     &                                                                                                                                                        & CS Guidance for Business                              & STEP 8.  Be prepared to react (Establish a formal   cyber incident management team who have been trained in and are following a   documented plan, which is tested at least annually.)                                                                                                                                                                                                                                                                                                                                                                                                                                                                                                                                                                                                                                                                                                                                                                                                                                                                                                                                                                                     &                                 \\ \cline{3-4}
                     &                                                                                                                                                        & OES Incident Notification Form                        & Operators of   essential services have to notify the CSIRT of an incident by using the   "Operators of Essential Services: Incident Notification Form"                                                                                                                                                                                                                                                                                                                                                                                                                                                                                                                                                                                                                                                                                                                                                                                                                                                                                                                                                                                                                     &                                 \\ \cline{3-4}
                     &                                                                                                                                                        & Incident Report Form                                  & Digital   service providers have to notify the CSIRT of an incident by by using the   "Digital Service Providers: Incident Notification Form"                                                                                                                                                                                                                                                                                                                                                                                                                                                                                                                                                                                                                                                                                                                                                                                                                                                                                                                                                                                                                              &                                 \\ \cline{3-4}
                     &                                                                                                                                                        & CSIRT-IE                                              & The government   has a unit (CSIRT, CERT, CIRT, etc.) that is specialised in national-level   cyber incident detection and response. The CSIRT-IE is responsible of the   incident response for national cyber security incidents at the government and   national levels.      &        

\\ \hline

\multicolumn{4}{|c|}{\textbf{\textit{Maturity Level Result}}}  & \textit{\textbf{3.1}}  
                     
\\ \hline

\end{longtable}
}

%% file: lessons_learned.tex
\section{Best Practices and Recommendations}

\subsection{Capability Assessment – Best Practices Observed}

The questions of the capability maturity assessment that received the highest scores, together with the related evidence, are reported below.

\subsubsection{Skills and Competency Development}

\textit{Question 6:} To what extent is information security management training developed and disseminated?

\begin{itemize}
  \item \textbf{National CS Strategy: }Skillnet Ireland launched their Cyber Security Skills Initiative to deliver a broad programme of initiatives in the field. Government has launched Future Jobs Ireland, a multiannual framework for skills and enterprise development, including the technology sector. Ireland offers a significant number of courses in cyber security, with at least 8 Masters level courses now on offer.
  \item\textbf{Operators of Essential Service:} All users are informed and trained on cyber security policies and relevant procedures, with periodic updates. (Detail standards)
\end{itemize}

\subsubsection{Cybersecurity threats}

\textit{Question 13:} What information about threats is available at the country level?

\begin{itemize}
  \item \textbf{IS Management:} "Appendix D offers some potential information sources to stay abreast of the latest threats, vulnerabilities, and potential security control options. This includes:
  \subitem - ENISA (Annual) Threat Landscape (ETL) Report provides an overview of threats, together with current and emerging trends.
  \subitem - ENISA Advisory Info Notes provides the information in relation to the latest threats and vulnerabilities."
  \item \textbf{CS Guidance for Business:} This guidance provides clear steps which organisations should include to understand who might want to attack them, why, and how they might go about carrying out such an attack in order to allow organisations to focus the efforts on how to respond to the most likely threats. For example, it recommends to establish a Cyber Threat Intelligence (CTI) capability which enables organisations to identify (through intelligence sources/feeds) and understand the top 5-10 threat actors and likely attack scenarios (these are your key cyber risks) and record them in a risk register. Please, see section 3. Understand the Threats.
  \item \textbf{Incident Report Form:} This incident report form includes a section of lessons learned based on new threats identified.
  \item \textbf{National CS Strategy:} 
  \subitem - Expanding the current Threat Sharing Group (TSG) which was established in 2017. The TSG acts both as a forum for critical national infrastructure operators, and a means for State Actors (including Gardaí and Defence Forces) to share information with these operators and to engage with cybersecurity professionals.
  \subitem - The NCSC has developed a threat intelligence database that is being used to assist Agencies and Departments in protecting their networks. 
  \subitem - The CSIRT has a proactive position that includes the deployment and use of MISPs (Malware Information Sharing Platform) to share threat intelligence directly with Critical National Infrastructure Providers, and the evolution and use of a series of tools to identify, parse and analyse open-source intelligence (OSINT). The CSIRT has also developed, tested, and deployed the ‘Sensor’ platform, now operational on the infrastructure of a number of Government Departments, to detect and warn of certain types of threat."
  \item \textbf{The National Cyber Security Centre (NSCS):}"The National Cyber Security Centre (NCSC) is regularly publishing Alerts \& Advisories on cyber security issues that may affect Ireland (https://www.ncsc.gov.ie/news/).
  
  \item \textbf{The European Union Agency for Cybersecurity (ENISA):} ENISA is the Union’s agency dedicated to achieving a high common level of cybersecurity across Europe. ENISA provides the reports EU on threats as below:
  \subitem - ENISA (Annual) Threat Landscape (ETL) Report provides an overview of threats, together with current and emerging trends. For example,  ENISA Threat Landscape 2020 - List of top 15 threats.

  \subitem- ENISA Advisory Info Notes provides the information in relation to the latest threats and vulnerabilities." Operators of Essential Service	Appendix B outlines the best practices and informative references recommended by the government for Operators of Essential Service in order to manage cyber threat information.

\end{itemize}

\subsubsection{Cybersecurity Threat Actors}

\textit{Question 14:} What information about threats actors is available at the country level?\\

\begin{itemize}
  \item \textbf{CS Guidance for Business:} "This guidance provides clear steps which organisations should follow to defend their business against threat actors. Please, refer to the following sections:\\

  \subitem 3. Understand the threats: Threat actors (cyber criminals, malicious insiders, etc.) vary in capability and sophistication, whilst also constantly changing depending on the value of the prize they seek to exploit.\\

  \subitem 5. Focus on education and awareness Establish an education and awareness programme, ensuring all of your employees, contractors and third parties can identify a cyberattack and are aware of the role they play in defending your business against threat actors.

  \subitem 7. Be able to detect an attack: Threat actors are many and sophisticated, and dedicated attackers have a high chance of breaching organisations' defences given enough time and persistence."
  \item \textbf{National CS Strategy: }This document identifies key threat actors such as those state-sponsored entities, usually military or security organisations, seeking to use network and information systems to conduct operations ranging from the exfiltration of data to the destruction of physical infrastructure. These threat actors, usually referred to as ‘advanced persistent threats’ (or APTs) have been shown to be involved in attacks across a wide range of sectors, but with a particular focus on Government IT systems, telecommunications networks, financial services, and technology companies.
  
  \item \textbf{The National Cyber Security Centre (NSCS): } "The National Cyber Security Centre (NCSC) is regularly publishing Alerts \& Advisories on cyber security issues that may affect Ireland (https://www.ncsc.gov.ie/news/).
			
   \item \textbf{The European Union Agency for Cybersecurity (ENISA):} "ENISA is the Union’s agency dedicated to achieving a high common level of cybersecurity across Europe. ENISA provides the reports EU on threats as below:
    \subitem - ENISA (Annual) Threat Landscape (ETL) Report provides an overview of threats, together with current and emerging trends. For example, ENISA Threat Landscape 2020 - List of top 15 threats.\\

    \subitem - ENISA Advisory Info Notes provides the information in relation to the latest threats and vulnerabilities."\\
\end{itemize}

\subsection{Sectorial best practices}

The questions of the qualitative findings for the sectorial best practices, together with the related evidence, are reported below.

What are the government cybersecurity guidelines/notes for the digital service providers?\\

\begin{itemize}
  \item \textbf{Digital Service Providers:} The Information Note for Digital Service Providers assists Digital Service Providers (i.e., Online marketplaces, Online search engines, and Cloud computing services) in understanding their obligations in relation to compliance with the NIS Directive.
\end{itemize}

What are the government cybersecurity guidelines/notes for operators of essential services?

\begin{itemize}
  \item \textbf{Operators of Essential Services:} The NIS Compliance Guidelines for Operators of Essential Service (OES)  (including digital infrastructure) establishes a set of Guidelines designed to assist OES in meeting their network and information system security and incident reporting requirements under Regulations 17 and 18 of the NIS Regulations.
\end{itemize}

\subsection{Recommendations based on the main findings}

\subsubsection{Strategy}

The National Cybersecurity Strategy of Ireland (2019) of Ireland should be updated in accordance with current and potential security threats and the recent lessons learnt. Such strategy has a foundation and initial direction for security activities. It contains a list of actions with times and leads but it is does not contain the detailed activities and it is not up-to-date. There are guidelines which focus on protection of  sensitive data and information and on establishing a set of perimeter barriers, and that begin to consider business/IT strategies and risk appetite.

\subsubsection{Implementation guidance}

The guidance for implementing the national strategy should be enhanced to better assist public and private companies in achieving their security objectives (e.g., improve the resilience and security of public sector IT systems). Appendix 1 of the National Cybersecurity Strategy of Ireland (2019) contains the list of actions for implementing the cybersecurity strategy but they are not very detailed and are not up-to-date. However, the NIS Compliance Guidelines for Operators of Essential Service (OES) establishes a set of Guidelines designed to assist OES in meeting their network and information system security and incident reporting requirements under Regulations 17 and 18 of the NIS Regulations. Moreover, the Information Note for Digital Service Providers assists Digital Service Providers in understanding their obligations in relation to compliance with the NIS Directive. Finally, the 12 Steps to Cyber Security Guidance on Cyber Security for Irish Business is intended to be used by businesses as a suggested activity plan which may be undertaken on a month-by-month basis over a suggested 12 month period to improve cyber resilience.

\subsubsection{Policy and Regulation}

Although some policies, regulations and national acts have been developed by the government of Ireland (e.g., NIS Compliance Guidelines for Operators of Essential Service and Information Note for Digital Service Providers), more comprehensive information security laws are required.  These policies and regulations should be aligned with changes in strategic priorities, technology trends, and security risks. 


\subsubsection{Threat Profiling}
Although, information about threats is available on the NCSC site, the current threats should be widely shared with all the  different subjects affected by cybersecurity issues (e.g., public sector, private sector, and citizens). Moreover, improvements in the identification of threats and threat actors could be made based on changes in the risk landscape and from lessons learned from previous information security incidents at the national level and internationally.

\subsubsection{Skills and Awareness}

Ireland should increase the general level of the skills and awareness of individuals around strong cybersecurity best practices and support them through information and training. Growing skills, and competencies as well as providing structure, governance, technical expertise and resources to assist with the response and recovery is one of the critical recommendations of public organisations such as HSE~\cite{web:hse2021report}. This must include structured educational programmes for delivering knowledge and skills training and scenario-based exercises to all relevant stakeholders.

%% file: conclusion.tex
\section{Summary and Outlook}
This report has presented the principles and characteristics of an instrument forn assessing the cybersecurity maturity  of the country level in order to help governments ensuring optimum response capability to cybersecurity threats and attacks and management of critical resources to prevent them.

The instrument has been used to evaluate the cybersecurity maturity of Ireland in four main categories, namely: Governance, Technical Security, Security Data Administration and Business Continuity Management.
The numerical results have been presented as well as the evidence that has led to such results. Also, some best practices and recommendations have been provided.

We plan to use the instrument to evaluate the cybersecurity maturity of other countries to make a comparison and derive the overall best practices that can help a country to implement the most effective cybersecurity strategies, policies and implementation actions to prevent and fight it. We envision that this work will influence policy makers to improve the planning, management and implementation of cybersecurity strategies.

%% file: bios.tex
\section{Additional Information}

\subsection{Notes on contributors}

\begin{itemize}

    \item [] \textbf{Marco Alfano}
    
    \textit{\textbf{Marco Alfano}} is a Senior Researcher at the Innovation Value Institute (IVI), Maynooth University, and leader of the IVI Digital Health research cluster. He is also affiliated with Lero, the SFI Research Centre for Software, and receives SFI funding for his research. He is currently working on responsible use of AI in health and well-being by facilitating person/patient empowerment and seamless communication within the healthcare system (http://cohealth.ivi.ie/). His research interests include Responsible AI, Digital Health Transformation, Patient Empowerment, Human-machine communication, Data analytics, Semantic Web, Smart cities, Cybersecurity, and Open Data/Big Data. He has authored more than fifty peer reviewed articles for journals, books, and conferences. He has participated in several European projects and has received grants from international bodies, such as the European Union (under the FP7 and H2020 framework programs), and national bodies, such as Science Foundation Ireland, Enterprise Ireland, and the National Research Council of Italy.
    
    \item [] \textbf{Viviana Bastidas}

    \textit{\textbf{Viviana Bastidas}} is a Research Associate at the Cambridge Centre for Smart Infrastructure and Construction (CSIC), University of Cambridge, UK. She works on the Digital Cities for Change (DC2) project which aims to develop a Competency Framework focusing on urban planning and responsible innovation. She is a research collaborator of IVI and the Enterprise Architecture and Formal Modelling (EAFM) research group, at Maynooth University. Her career interests are in the intersection of business and Information Technologies in the context of socio-technical design to support urban governance and planning for smart cities. Viviana’s research has been published in conferences and journals in the domains of Business and Information Systems Engineering, Knowledge-Economy, Digital Transformation, the Internet of Things, and System Science. 
    
    \item [] \textbf{Paul Heynen}

    \textit{\textbf{Paul Heynen}} has been working with IVI as operations manager since its formation in 2006. Paul is responsible for the non-research business operations functions in IVI including community, member and partner engement, financial administration, liaison with central university administrative functions, and on-going management and development of IVI-specific IT systems and platforms. Prior to joining IVI, Paul worked in the semiconductor industry for 11 years in various engineering and management roles, most recently as a product manager for an advanced process control software and analytics start-up.
    
    \item [] \textbf{Markus Helfert}
    
    \textit{\textbf{Markus Helfert}} is the Director of IVI and Director of Empower – the SFI funded Programme on Data Governance. He is also Professor of Digital Service Innovation and Director of the Business Informatics Group at Maynooth University. He is a Principle Investigator at Lero – The Irish Software Research Centre and at the Adapt Research Centre. His research is centred on Digital Service Innovation, Smart Cities and IoT based Smart Environments and includes research areas such as Service Innovation, Intelligent Transportation Systems, Smart Services, Building Information Management, FinTech, Data Value, Enterprise Architecture, Technology Adoption, Analytics, Business Process Managem ent. Prof. Helfert is an expert in Data Governance Standards and is involved in European Standardisation initiatives. Markus Helfert has authored more than 200+ academic articles, journal and book contributions and has presented his work at international conferences. Helfert has received national and international grants from agencies such as European Union (FP7; H2020), Science Foundation Ireland and Enterprise Ireland, was project coordinator on EU projects, and is the Project coordinator of the H2020 Projects: PERFORM on Digital Retail
    
\end{itemize}

%% file: main.bbl
\begin{thebibliography}{}

\bibitem[Al-rimy et~al., 2018]{al2018ransomware}
Al-rimy, B. A.~S., Maarof, M.~A., and Shaid, S. Z.~M. (2018).
\newblock Ransomware threat success factors, taxonomy, and countermeasures: A
  survey and research directions.
\newblock {\em Computers \& Security}, 74:144--166.

\bibitem[Chigada and Madzinga, 2021]{chigada2021cyberattacks}
Chigada, J. and Madzinga, R. (2021).
\newblock Cyberattacks and threats during covid-19: A systematic literature
  review.
\newblock {\em South African Journal of Information Management}, 23(1):1--11.

\bibitem[Dedeke and Masterson, 2019]{dedeke2019contrasting}
Dedeke, A. and Masterson, K. (2019).
\newblock Contrasting cybersecurity implementation frameworks (cif) from three
  countries.
\newblock {\em Information \& Computer Security Journal}.

\bibitem[PwC, 2021]{web:hse2021report}
PwC (2021).
\newblock Conti cyberattack on the hse report.
\newblock
  https://www.hse.ie/eng/services/publications/conti-cyber-attack-on-the-hse-full-report.pdf.
\newblock Last accessed 08 February 2023.

\bibitem[Sabillon et~al., 2016]{sabillon2016national}
Sabillon, R., Cavaller, V., and Cano, J. (2016).
\newblock National cyber security strategies: global trends in cyberspace.
\newblock {\em International Journal of Computer Science and Software
  Engineering}, 5(5):67.

\bibitem[Sadik et~al., 2020]{sadik2020toward}
Sadik, S., Ahmed, M., Sikos, L.~F., and Islam, A. (2020).
\newblock Toward a sustainable cybersecurity ecosystem.
\newblock {\em Computers}, 9(3):74.

\bibitem[Sarri et~al., 2020]{enisa2020frame}
Sarri, A., Kyranoudi, P., Thirriot, A., Charelli, F., and Dominique, Y. (2020).
\newblock National capabilities assessment framework.
\newblock {\em The European Union Agency for Cybersecurity (ENISA)}.

\end{thebibliography}
